\documentclass[prd,amsfonts,aps,nofootinbib,notitlepage,11pt,superscriptaddress]{revtex4-1}

\pdfoutput=1
\usepackage{amsmath,amssymb}
\usepackage{graphicx}
\usepackage[utf8]{inputenc}
\usepackage{cancel}
\usepackage[colorlinks=true]{hyperref}
\usepackage{color}
\usepackage{array,multirow}
\usepackage{subcaption}
\usepackage{float}
\usepackage{comment}

\newcommand{\UPTC}{Escuela de Física, Universidad Pedagógica y Tecnológica de Colombia,\\
Avenida Central del Norte \# 39-115, Tunja, Colombia}
\newcommand{\UdeA}{Instituto de Física, Universidad de Antioquia,\\Calle 70 \# 52-21, Apartado Aéreo 1226, Medellín, Colombia}

\begin{document}
\title{A minimal model of fermion FIMP dark matter}
\author{Carlos E. Yaguna}
\affiliation{\UPTC}
\author{\'Oscar Zapata}
\affiliation{\UdeA}

\begin{abstract}
 We investigate a simple extension of the standard model (SM) in which the dark matter consists of a feebly interacting  fermion (FIMP),  charged under a new $Z_4$ symmetry, that is produced in the early Universe by the freeze-in mechanism. The only other new particle included in the model is  a singlet scalar, also charged  under the $Z_4$, which couples to  the fermion via Yukawa interactions and to the SM Higgs. The model is truly minimal,  as it admits just five free parameters: two masses and three dimensionless couplings. Depending on their values, the  freeze-in mechanism can be realized in different ways, each characterized by its own production processes. For all of them, we numerically study the relic density as a function of the free parameters of the model  and determine the regions consistent with the dark matter constraint. Our results show that  this scenario is viable over a wide range of couplings and dark matter masses. This model, therefore, not only offers a novel solution to the dark matter problem, but it also provides a minimal realization of freeze-in for fermion dark matter.  
  \end{abstract}

\maketitle
\section{Introduction}
The observed dark matter abundance \cite{WMAP:2012nax, Aghanim:2018eyx} could be explained by a feebly interacting massive particle (FIMP)~\cite{Bernal:2017kxu,Agrawal:2021dbo} that was produced  via the freeze-in mechanism. Freeze-in assumes that the dark matter particle interacts so weakly that it never reached equilibrium with the thermal plasma in the early Universe~\cite{Hall:2009bx}, resulting in a simple and very predictive framework. Indeed, once the particle physics model is known, the FIMP relic density can be calculated, within the standard cosmological model, without the need of any extra parameters or hypotheses  --just as for the conventional freeze-out~\cite{Steigman:2012nb}. It is such the attention that freeze-in is receiving lately in the literature that it has already been incorporated into the computational codes used by the dark matter community, including micrOMEGAs \cite{Belanger:2018ccd} and, more recently, DarkSUSY \cite{Bringmann:2021sth}.  FIMP dark matter may also be the reason why no signal has so far been observed in dark matter detectors~\cite{Bertone:2018krk}. 

Given that FIMPs are necessarily SM singlets, the minimal scenario that realizes freeze-in is the singlet scalar model \cite{Yaguna:2011qn}. In it, the dark matter particle is a real scalar ($S$)  odd under a new $Z_2$ symmetry that couples directly only to the Higgs field ($H$) --via the quartic interaction  $\lambda_{HS}S^2H^\dagger H$~\cite{Silveira:1985rk,McDonald:1993ex,Burgess:2000yq}. This model has only two free parameters --$\lambda_{HS}$ and the $S$ mass-- and is consistent with the observed dark matter density for  $\lambda_{HS}\sim 10^{-11}$. But, when  the dark matter is instead a fermion, the model cannot be so simple. Since  it is not possible to couple a dark matter fermion to the SM via renormalizable terms, the model would inevitably require extra fields. One possibility is to add a scalar, even under a $Z_2$ symmetry, that mixes with the Higgs, as done in the singlet fermionic model~\cite{Kim:2006af,Kim:2008pp, Lopez-Honorez:2012tov,Esch:2013rta}. Although consistent with freeze-in~\cite{Klasen:2013ypa}, this appealing scenario contains $7$ free parameters, significantly more than for scalar dark matter. It is natural to ask, therefore, whether there exists a simpler realization of freeze-in for fermion dark matter, one featuring a smaller number of free parameters. 

In this work, we study the DM model proposed in Refs.~\cite{Cai:2015zza,Yaguna:2021rds} but in light of the freeze-in mechanism, rather than the freeze-out considered there.  This minimal extension of the SM incorporates just two new particles, a Dirac fermion and a real scalar, both SM singlets though  charged under a new  $Z_4$ symmetry. Remarkably, only five free parameters --the two masses and three dimensionless couplings-- are allowed in this setup.   We show that,  depending on the relation between the masses and on the values of the couplings,  freeze-in can be realized in different ways within this model. For each of them, we compute the  relic density and determine the regions consistent with the dark matter constraint. Hence, we demonstrate that this attractive and viable scenario provides a minimal realization of freeze-in for fermion dark matter. 

The rest of the paper is organized as follows. In the next section the model is introduced and its free parameters are identified. Section \ref{sec:FIMPDM}  qualitatively describes FIMP dark matter  and presents the possible realizations of the freeze-in mechanism within this model. Our main results are obtained and explained in section \ref{sec:results}. There, we determine, for each realization of freeze-in, the regions of parameter space that are consistent with current data. In section \ref{sec:discussion} the $Z_4$ model of freeze-in is contrasted against related scenarios, and possible extensions are briefly discussed. Finally, we summarize our results and present our conclusions in section \ref{sec:conclusions}.

\section{The model}\label{sec:model}
We extend the SM  with a new $Z_4$ discrete symmetry and two additional fields, a fermion ($\psi$) and a real scalar ($S$), both singlets of the SM gauge group. Under the $Z_4$, all the SM particles are  singlets, whereas the new fields  transform as $S\to -S$ and $\psi\to i\psi$.  That is all. 

The most general Lagrangian symmetric under $SU(3)\times SU(2)\times U(1)\times Z_4$  contains the following new terms 
\begin{align}\label{eq:L}
 \mathcal{L}&=\,\,-\frac{1}{2}\mu_{S}^2S^2-\frac{1}{4}\lambda_{S}S^4 -\frac{1}{2}\lambda_{S H}|H|^2S^2-M_{\psi}\overline{\psi}\psi+\frac{1}{2}\left[y_s\overline{\psi^c}\psi + y_{p}\overline{\psi^c}\gamma_5\psi + \rm{h.c.}\right]S, 
 \end{align}
where $H=[0, (h+v_H)/\sqrt{2}]^T$, with $h$ the SM Higgs boson. Unlike the Higgs, the new scalar field $S$ does  not  acquire a vacuum expectation value. To guarantee that the tree-level potential remains bounded from below, we require $\lambda_S>0$ and $\lambda_{SH}+2\sqrt{\lambda_H\lambda_S}>0$, where $\lambda_{H}$ is the SM Higgs quartic coupling.

 The first two terms in the above Lagrangian involve only $S$ and correspond respectively to the usual mass term and quartic interaction, while the third term is the so-called Higgs portal interaction between $S$ and the SM Higgs. The  three subsequent terms involve $\psi$ and amount to a Dirac mass term plus two Yukawa interactions between $\psi$ and $S$. Due to the $Z_4$ invariance, these Yukawa terms feature a Majorana-like structure ($\overline{\psi^c}\psi$), rather than the more conventional Dirac one ($\overline{\psi}\psi$). Notice that while $S$ couples to the Higgs, $\psi$ does not couple to any SM fields --only to $S$.  

The free parameters of this model are just five ($\lambda_S$ is irrelevant to our discussion), which can be taken to be 
\begin{align}
    M_S, M_\psi, \lambda_{SH}, y_s, y_p, 
\end{align}
where the $S$ mass, $M_S$, is given by
\begin{align}
    M_{S}^2&=\mu_S^2+\frac{1}{2}\lambda_{SH}v_H^2.
\end{align}
For definiteness,  $y_s$ and $y_p$ will be taken to be real in the following. In addition,  due to the  analogous roles they play in the Lagrangian (\ref{eq:L}), we will usually  considered them jointly in our subsequent discussions.

It follows from Eq.~(\ref{eq:L}) that the fermion is automatically stable, whereas the scalar can be stable, if $M_S>2M_\psi$, or unstable, if  $M_S>2M_\psi$ --decaying into two fermions via the Yukawas. The model thus may accommodate one ($\psi$) or two ($\psi,S$) dark matter particles. In this work, we will consider \emph{both} possibilities while focusing our attention on the region of parameter space where the dark matter is feebly-interacting and is produced in the early Universe by the freeze-in mechanism. 

Recently, we investigated the above Lagrangian  but within the context of a two-component WIMP dark matter framework \cite{Yaguna:2021rds}, and  found that it gives rise to a consistent and testable scenario for the  dark matter. In Ref. \cite{Barman:2020jrf}, this $Z_4$ model was extended with a non-renormalizable term and studied within a non-standard cosmological model. The region of parameter space considered in these previous works, as well as the associated dark matter phenomenology, are, however, completely different from those we are going to investigate in this paper. The phenomenological implications when the scalar field gets a non-zero vacuum expectation value have been studied in Refs.~\cite{Finkbeiner:2007kk, Bell:2010qt, Bhattacharya:2018ljs}. 
The $Z_4$ as a stabilizing dark matter symmetry was  employed in Ref.~\cite{Belanger:2022qxt}, which features a different particle content --two scalars, a singlet and a doublet. A related model using a $Z_2$ symmetry instead of a $Z_4$  was analyzed in Ref.~\cite{Klasen:2013ypa} for freeze-in dark matter. Besides the $Z_2$ odd fermion, it includes a scalar that is even under the $Z_2$ and  mixes with the Higgs boson. Though viable, this $Z_2$ model contains two additional parameters in the scalar potential, so it is not as simple or predictive as the $Z_4$ we are proposing here. 

\section{FIMP Dark Matter}
\label{sec:FIMPDM}

We are interested in the case where the dark matter is, totally or partially,  explained via the freeze-in mechanism. The fundamental characteristic of freeze-in is that the dark matter particle interacts so feebly that is unable to ever reach thermal equilibrium in the early Universe --unlike for freeze-out. Thus, FIMPs are necessarily singlets of the SM gauge group and usually feature tiny couplings in their interactions with the SM particles. Through such interactions, FIMPs are slowly produced from the thermal plasma, their abundance increasing as the Universe expands, until it freezes-in at a temperature close to the FIMP mass \cite{Hall:2009bx}. Hence, it is the production, rather than the annihilation, processes that ultimately determine the FIMP relic density. 

In the $Z_4$ model we are studying, both $S$ and $\psi$ can be FIMPs. If the Yukawa interactions are suppressed, $y_{s,p}\ll 1$, $\psi$ will be a FIMP and, due to its stability, a dark matter particle. If, instead, it is the Higgs portal interaction that is suppressed, $\lambda_{HS}\ll1$, both $S$ and $\psi$ will be FIMPs. Indeed, once $S$ is feebly interacting, $\psi$ will automatically inherit that trait, as a result of its couplings. In this second scenario, $S$ may or not be a dark matter particle, contingent on its stability.  

\begin{figure}[t]
\centering
\includegraphics[scale=0.45]{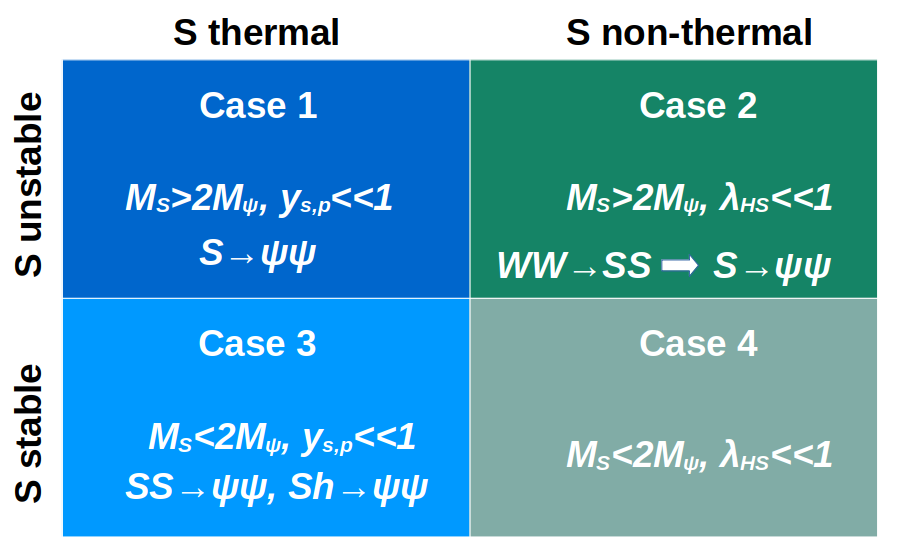}
\caption{The four possible realizations of freeze-in within the $Z_4$ model. In this diagram by "$S$ thermal" ("$S$ non-thermal") we mean that $S$ was able (unable) to thermalize with the plasma at some point in the early Universe.  In the first row, $\psi$ is the dark matter whereas in the second row $\psi$ and $S$ are both dark matter particles --they both contribute to the observed dark matter density. In case 4 the contribution from $\psi$ turns out to be entirely negligible, however.}
\label{fig:cases}
\end{figure}

Within freeze-in, the specific processes that generate the  FIMP dark matter abundance are model-dependent, being $1\to 2$ decays and $2\to 2$ scatterings the most common ones.  In this model, we have identified four distinctive ways through which the freeze-in mechanism can be realized, each featuring its own production mechanisms. These four \emph{cases}, which depend  on $S$ reaching or not thermal equilibrium and on its stability, are as follows:
\begin{itemize}
\item Case 1: $S$ is unstable and reaches thermal equilibrium in the early Universe. The production of dark matter, $\psi$, occurs via the decay $S\to \psi+\psi$. The Yukawa couplings are tiny in this case, $y_{s,p}\ll1$.

\item Case 2: $S$ is unstable and does not reach thermal equilibrium in the early Universe. Dark matter production is a two step process in this case. First, $S$ is pair-produced by scatterings of SM particles in the thermal plasma, e.g. $W^++W^-\to S+S$. Then, $S$ decays into dark matter particles --$S\to \psi+\psi$. The tiny coupling in this case is the Higgs portal, $\lambda_{HS}\ll 1$.

\item Case 3: $S$ is stable and reaches thermal equilibrium in the early Universe. $S$ and $\psi$ are both dark matter particles, with the former being a WIMP while the latter a FIMP. $S$ undergoes the usual freeze-out while $\psi$ is produced from $S$ via $S+h\to \psi+\psi$ and $S+S\to \psi+\bar \psi$. This production mechanism, unlike the previous ones, is specific to multi-component dark matter scenarios, and will be dubbed \emph{FIMP from WIMP} in the following\footnote{See Refs. \cite{DuttaBanik:2016jzv, Bhattacharya:2021rwh, Costa:2022lpy} for two-component FIMP-WIMP scenarios.}. The Yukawa couplings are again tiny in this case, $y_{s,p}\ll1$.

\item Case 4: $S$ is stable and does not reach thermal equilibrium. In this case the fermion cannot be produced in any significant amount, so $S$ is the dark matter FIMP and the model effectively reduces to the already known freeze-in realization of the singlet scalar \cite{Yaguna:2011qn, Belanger:2018ccd, Bringmann:2021sth}. For this reason, we will not  consider this possibility any further. 

\end{itemize}
To emphasize that these cases constitute really independent situations rather than just different ways of generating the dark matter, they are displayed in matrix form in figure \ref{fig:cases}.  In the following section, we will analyze in detail cases 1 to 3 so as to determine the regions of parameter space where the observed dark matter density can be explained.

\section{Results}
\label{sec:results}
In this section we present our main results. For each case, we will first identify the relevant dark matter production processes and study quantitatively the dependence of the relic density  on the free parameters of the model. Then, the viable regions will be obtained and illustrated. For our numerical findings we will rely on micrOMEGAs, which since its version 4.1~\cite{Belanger:2014vza} was extended to multicomponent dark matter scenarios (case 3), and since its version 5.0 \cite{Belanger:2018ccd} incorporated freeze-in. In micrOMEGAs, the different processes that contribute to the dark matter density are automatically taken into account, and the corresponding Boltzmann equations  are numerically and accurately solved. To implement the  $Z_4$ model into micrOMEGAs, we used LanHEP~\cite{Semenov:2014rea}. 

\subsection{Case 1}
This case is realized for $M_S>2M_\psi$,  $y_p,y_s\ll 1$ and a $\lambda_{SH}$ large enough to allow $S$ to reach thermal equilibrium at high temperatures. $\psi$,  the dark matter particle, is feebly interacting and is dominantly  produced via $S$ decays.  These decays  can happen while $S$ is  in thermal equilibrium (the freeze-in contribution) or after $S$ freezes-out (the SuperWIMP contribution). The total dark matter density will then be the sum of these two contributions, which we now discuss separately.

\subsubsection{The freeze-in contribution}
Since $\psi$ has a direct coupling to $S$ (via the Yukawa couplings), it will be produced  by the decay of $S$ while it is  in equilibrium with the thermal bath. As long as $M_S>2M_\psi$,  the scalar decays into a pair $\psi\psi$ or $\psi^c\psi^c$, with a total rate given by 
\begin{align}
    \Gamma(S\to2\psi)\equiv\Gamma(S\to \psi\psi+\psi^c\psi^c)&=\frac{M_S}{8\pi}\left[|y_s|^2\left(1-\frac{4M_\psi^2}{M_S^2}\right)^{3/2} + |y_p|^2\left(1-\frac{4M_\psi^2}{M_S^2}\right)^{1/2}\right]\label{eq:drate}\\
    &\approx \frac{M_S}{8\pi}\left[|y_s|^2 + |y_p|^2\right],
\end{align}
with the last expression being valid far from the threshold. Equivalently, 
the $S$ lifetime, $\tau_S$, can be written as
\begin{align}
    \tau_S\simeq 10^{-9}\,\text{sec}\,\left(\frac{100\,\mathrm{GeV}}{M_S}\right)\left(\frac{10^{-8}}{|y_{s,p}|}\right)^2.
\end{align}

The $\psi$ yield, $Y_{\psi}(T)=n_{\psi}(T)/s(T)$, is computed by solving the following Boltzmann equation~\cite{Hall:2009bx}
\begin{equation}
	s\, T\,\frac{d Y_{\psi}}{dT} \; = \; -\frac{\gamma_{S\to2\psi}(T)}{H(T)}\,,\label{BE}
\end{equation}
where $s$ is the entropy density of the Universe, $H(T)$ is the expansion rate of the Universe at a given temperature and $\gamma_{S\to2\psi}(T)$ is the thermal averaged FIMP production rate, 
\begin{align}
    \gamma_{S\to2\psi}(T)= g_S\frac{M_{S}^{2}\,T} {2\,\pi^{2}}\,K_{1}\left(M_{S}/T\right)\,\Gamma(S\to2\psi),
\end{align}
with $K_{1}(x)$  the Bessel function of the second kind. It follows that
\begin{align}
    Y_\psi&\equiv Y_\psi(T_0)
    =g_S\frac{M_{S}^{2}} {\pi^{2}}\,\Gamma(S\to2\psi)\int_{T_0}^{T_R}\frac{dT\,K_{1}\left(M_{S}/T\right)}{H(T)s(T)}\nonumber\\
    &\approx0.685\frac{g_S\,M_P} {M_S^2}\,\Gamma(S\to2\psi)\left(\frac{1+1/3g_s'}{g_\rho^{1/2}g_s}\right)_{T=M_S/3},
\end{align}
where $s(T)=2\pi^2g_s T^{3}/45$, $H(T)=1.66\sqrt{g_\rho} T^{2}/M_{P}$ and $K_{1}(x)\sim1/x$ for $x\ll 1$.  For high temperatures, $T> M_{S}$, we obtain
\begin{equation}
	\frac{d Y_{\psi}}{dT} \; \approx \; - \,5\times 10^{3}\,\text{GeV}^{3}\,\left(\frac{M_{S}}{1\,\text{TeV}}\right)^{2}\,\left(\frac{y_{s,p}}{10^{-8}}\right)^{2}\,T^{-4}\,.\label{dYapprox}
\end{equation}
Therefore, on the one hand we have that at $T> M_{S}$ the yield always scales as the square of the scalar mass and of the  Yukawa coupling $y_{s,p}$. On the other hand, at $T\lesssim M_{S}$ the scalar particle abundance becomes Boltzmann suppressed  and  the production of  dark matter is no longer efficient. As a result we have
\begin{equation}
	Y_{\psi}\left(T\lesssim M_{S} \right)\;\approx\;  10^{-4}\,\left(\frac{1~\text{TeV}}{M_{S}}\right)\,\left(\frac{y_{s,p}}{10^{-8}}\right)^{2}\,.\label{Yapprox}
\end{equation}

\begin{figure}[t]
\centering
\includegraphics[scale=0.8]{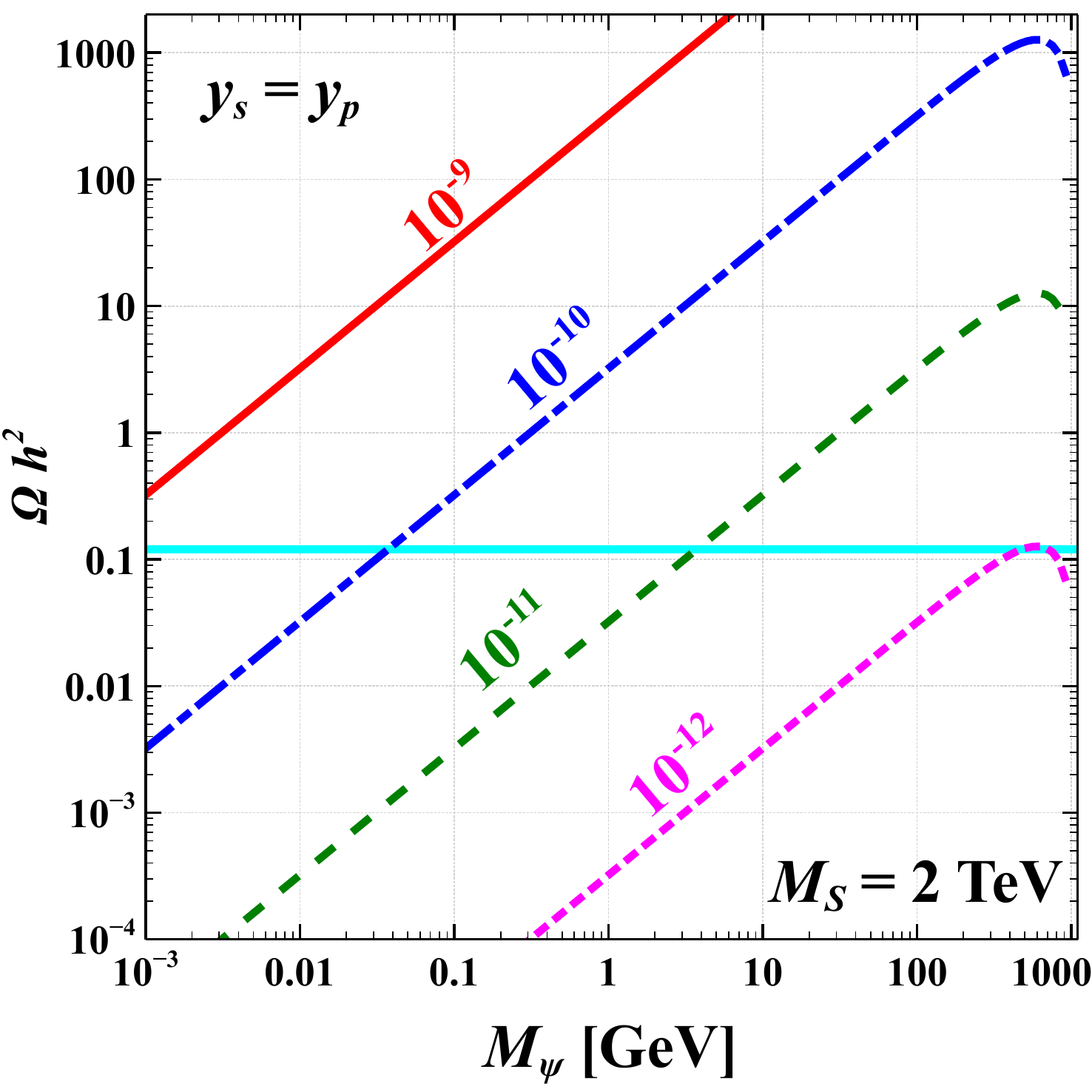}
\caption{The freeze-in relic density as a function of the dark matter mass for different values of the Yukawa couplings --from  $10^{-9}$ (top) to $10^{-12}$ (bottom). For this plot the scalar mass was set to $2$ TeV and $y_s=y_p$ was assumed. For illustration, the allowed range of the relic density is shown as a horizontal band.}
\label{fig:reliccase1}
\end{figure}

The relic density of dark matter, $\Omega_{\psi}h^{2}$, is related to the asymptotic value of $Y_{\psi}$ at low temperatures by
\begin{equation}
	\Omega_{\psi}\,h^{2}\;=\; 2.744\times 10^{8}\, \frac{M_{\psi}}{\text{GeV}}\,Y_{\psi}(T_{0})\,,\label{OmegaN1}
\end{equation}
where $T_{0}=2.752$ K is the present day CMB temperature. It is this quantity that should be compared with the observed dark matter density as measured by WMAP \cite{WMAP:2012nax} and PLANCK \cite{Aghanim:2018eyx}. For dark matter production via the freeze-in mechanism, the $\psi$ relic abundance  can be estimated as 
\begin{align}
  \Omega_{\psi}\,h^{2}&\approx 0.3\left(\frac{M_\psi}{0.1\,\mbox{GeV}}\right)\left(\frac{1\,\mbox{TeV}}{M_S}\right)\left(\frac{y_{s,p}}{10^{-10}}\right)^2,
\label{eq:rd}
\end{align}
where we used equations~(\ref{Yapprox}) and (\ref{OmegaN1}). Notice that this expression has the expected dependence on $M_S$, $y_{s,p}$ and $M_\psi$. 

\begin{figure}[t]
\centering
\includegraphics[scale=0.8]{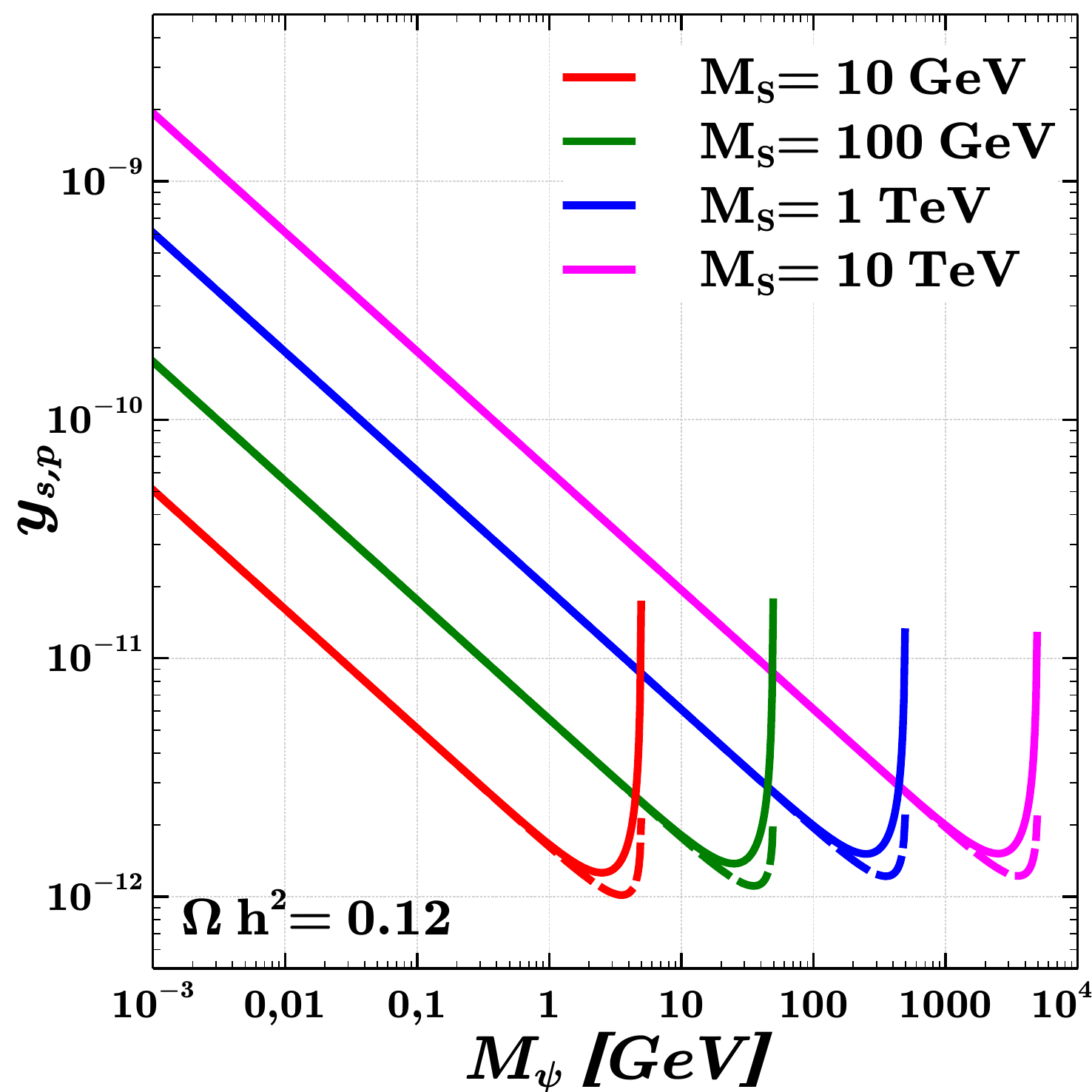}
\caption{The regions consistent with the relic density for case 1. Only the freeze-in contribution is shown. The solid (dashed) lines correspond to the value of $y_s$ ($y_p$) assuming that $y_p=0$ ($y_s=0$). }
\label{fig:pscase1}
\end{figure}

Figure \ref{fig:reliccase1} displays the freeze-in relic density as a function of the dark matter mass ($M_\psi$) for different values of the Yukawa couplings, from $10^{-9}$ (solid) to $10^{-12}$ (dotted). Let us stress that the results shown in this and all the following figures were obtained with micrOMEGAs and not with the analytical expressions obtained in the text, which serve instead as a check and illustrate the functional dependence with the different parameters.  From figure \ref{fig:reliccase1} we can see that equation (\ref{eq:rd}) indeed gives an excellent approximation to the relic density. The only exception is very close to the decay threshold, where the kinematic factors of $(1-4M_\psi^2/M_S^2)^n$ from the rate --see Eq.~(\ref{eq:drate})-- become important, changing the behaviour of the relic density with $M_\psi$. For this figure, we took  $y_p=y_s$ and $M_S=2$ TeV, which fixes the maximum allowed value of $M_\psi$ to $1$ TeV. The only other parameter of the model is $\lambda_{HS}$, which was assumed to be large enough to ensure that $S$ reaches thermal equilibrium --its precise value being inconsequential for the freeze-in relic density.  For comparison, the allowed value of the relic density~\cite{WMAP:2012nax, Aghanim:2018eyx} is displayed as a horizontal (cyan) band.  As expected, small values of the Yukawa couplings are required to be consistent with the observed dark matter density.

Having computed the  relic density, we can now proceed to obtain the viable regions for this case --that is, the regions of the parameter space where the relic density is consistent with the valued determined by Planck. Figure \ref{fig:pscase1} shows these viable regions for different values of $M_S$, from $10$ GeV (bottom line) to $10$ TeV (top line). In this figure we considered two different possibilities for the Yukawa couplings: $i)$ $y_s\neq 0,y_p=0$ (solid lines); $ii)$ $y_s= 0,y_p\neq0$ (dashed lines). Notice that these two possibilities --scalar and pseudoscalar interactions, respectively-- differ only very close to the decay threshold, as expected from equations (\ref{eq:rd}) and (\ref{eq:drate}). The value of the Yukawa couplings required to explain the relic density via freeze-in  thus lies approximately between $10^{-12}$ and $10^{-9}$, depending on the ratio $M_\psi/M_S$. Such small Yukawas guarantee that $\psi$ does not reach thermal equilibrium in the early Universe, as assumed in the freeze-in mechanism. To remain  consistent with \emph{warm} dark matter bounds, the range of $M_\psi$ can be  extended down to the $\sim 10$ keV range~\cite{DEramo:2020gpr, Decant:2021mhj} (not shown in figure \ref{fig:pscase1}).  The corresponding value of the couplings can be easily obtained by  extrapolating from the figure, or directly from equation (\ref{eq:rd}).

\subsubsection{The superWIMP contribution}
Since the Yukawa couplings are  tiny in this case, $S$ is meta-stable and undergoes a freeze-out before decaying and disappearing completely from the hot plasma in the early Universe. This late decay, $S\to \psi\psi$, takes place at $T\lesssim M_S/20$ and constitutes an additional source of dark matter called the superWIMP contribution ($\Omega^{sWIMP}_{\psi}$)~\cite{Feng:2003uy, Feng:2003xh}. Notice that all of the $S$ particles remaining after freeze-out will end up being converted into dark matter through the decay. Thus,
\begin{equation}
	\Omega^{sWIMP}_{\psi}=\frac{M_{\psi}}{M_{S}}\, 2\widetilde{\Omega}_{S}^{fo},
\label{swimp}
\end{equation} 
where $\widetilde{\Omega}_{S}^{fo}$ is the relic abundance of $S$, obtained via the usual \emph{freeze-out} mechanism, if it were stable. 
The crucial point for us is that while $\widetilde{\Omega}_{S}^{fo}$ depends on $\lambda_{HS}$ but not on $y_{s,p}$,  it is the other way around for the freeze-in contribution. Consequently, we can usually find a value of $\lambda_{HS}$ such that $\Omega^{sWIMP}_{\psi}\ll \Omega_{DM}$ and freeze-in accounts for most of the dark matter. Taking $\lambda_{HS}=1$, for instance, ensures that $\widetilde\Omega_S^{fo}<\Omega_{DM}$ for all $M_S\lesssim 7$ TeV.

\subsection{Case 2}

This case features  $M_S>2M_\psi$ and  $\lambda_{SH}\lesssim 10^{-6}$, which results in both $S$ and $\psi$ being feebly interacting, but since $S$ is unstable, only $\psi$ contributes to the dark matter. The production of dark matter is now a two-step process. First,  $S$ is slowly pair-produced via scattering of SM particles in the plasma (at $T\gtrsim M_S$), just as in the singlet scalar model \cite{Yaguna:2011qn}. Then, once $S$ has been  produced, it will necessarily (and immediately) decay into fermions (the dark matter) via the Yukawa interactions. A crucial difference with respect to case 1 is that the precise value of the Yukawa couplings does not affect the dark matter density --two $\psi$'s are produced for each $S$ irrespective of $y_{s,p}$. In fact,
\begin{equation}
\label{eq:relics}
\Omega=\frac{M_\psi}{M_S}\,2\widetilde\Omega_S^{fi},
\end{equation}
where the factor $2$ comes from $S\to \psi\psi$, and $\widetilde\Omega_S^{fi}$ is the $S$ \emph{freeze-in} relic density if it were stable --i.e. the one computed in the singlet scalar model. Consequently, the relevant parameter space consists of $M_S$, $M_\psi$ and $\lambda_{HS}$.

\begin{figure}[t]
\centering
\includegraphics[scale=0.8]{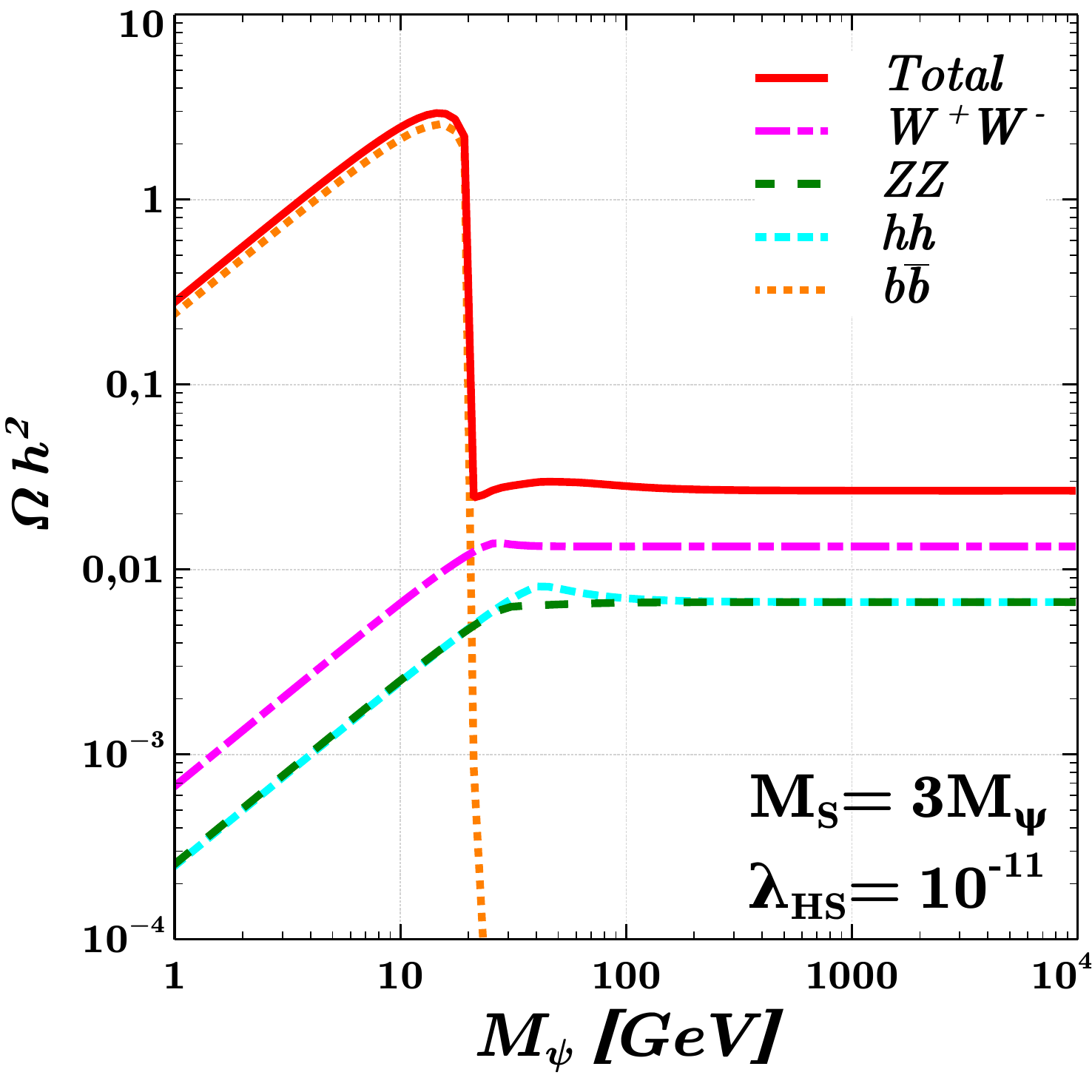}
\caption{The contributions of the different initial states to the relic density as a funcion of $M_\psi$. In this case, $S$ is pair-produced from SM particles and subsequently decays into $\psi$.}
\label{fig:relic2}
\end{figure}

Figure \ref{fig:relic2} shows, as a function of $M_\psi$, the contributions  to the dark matter density from different initial states: $b\bar b$ (dotted orange line), $W^+W^-$ (dash-dotted magenta line), $ZZ$ (dashed green line), and $hh$ (dotted cyan line). The total relic density is the solid (red) line.  For this figure we set $M_S=3M_\psi$ and $\lambda_{HS}=10^{-11}$. Below the Higgs resonance ($M_S<M_h/2$ or equivalently $M_\psi<M_h/6$) the production is entirely dominated by the $b\bar b$ initial state --dark matter is generated via $b+\bar b\to S+S$ (a Higgs-mediated process) followed by $S\to \psi+\psi$. Above the Higgs resonance, instead, the $W^+W^-$, $ZZ$ and $hh$ initial states all give sizable contributions to the relic density. These results are consistent with those already known from the singlet scalar model. Indeed, the drop in the value of $\Omega h^2$ at the Higgs-resonance as well as its constant behaviour at high dark matter masses are well-known features of this model.  Since $M_\psi/M_S$ is constant in this figure, $\Omega$ is simply proportional to $\widetilde\Omega_S^{fi}$, in agreement with equation (\ref{eq:relics}) --see e.g. Refs.~\cite{Yaguna:2011qn, Belanger:2018ccd, Bringmann:2021sth}. 

\begin{figure}[t]
\centering
\includegraphics[scale=0.53]{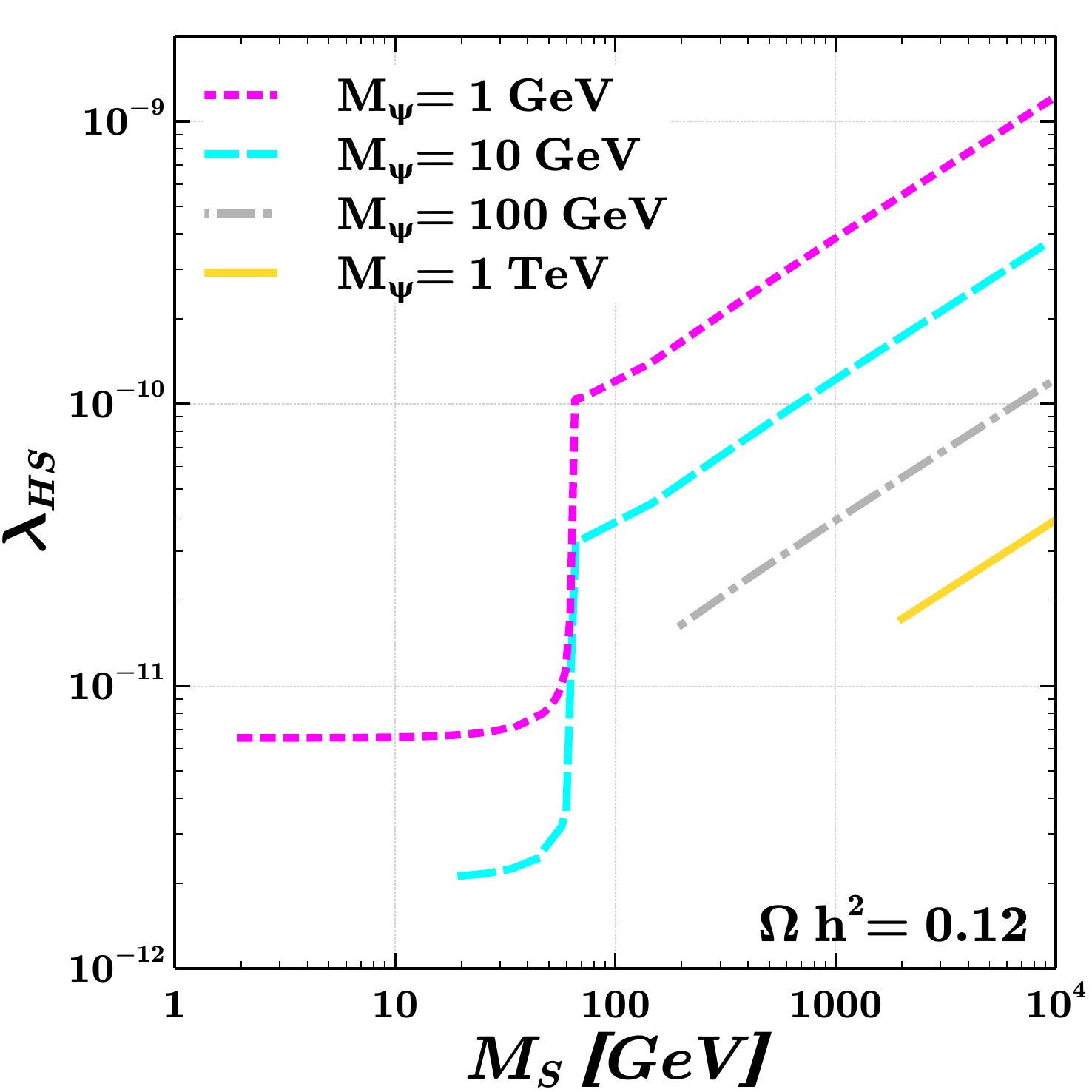}\hspace{4mm}\includegraphics[scale=0.53]{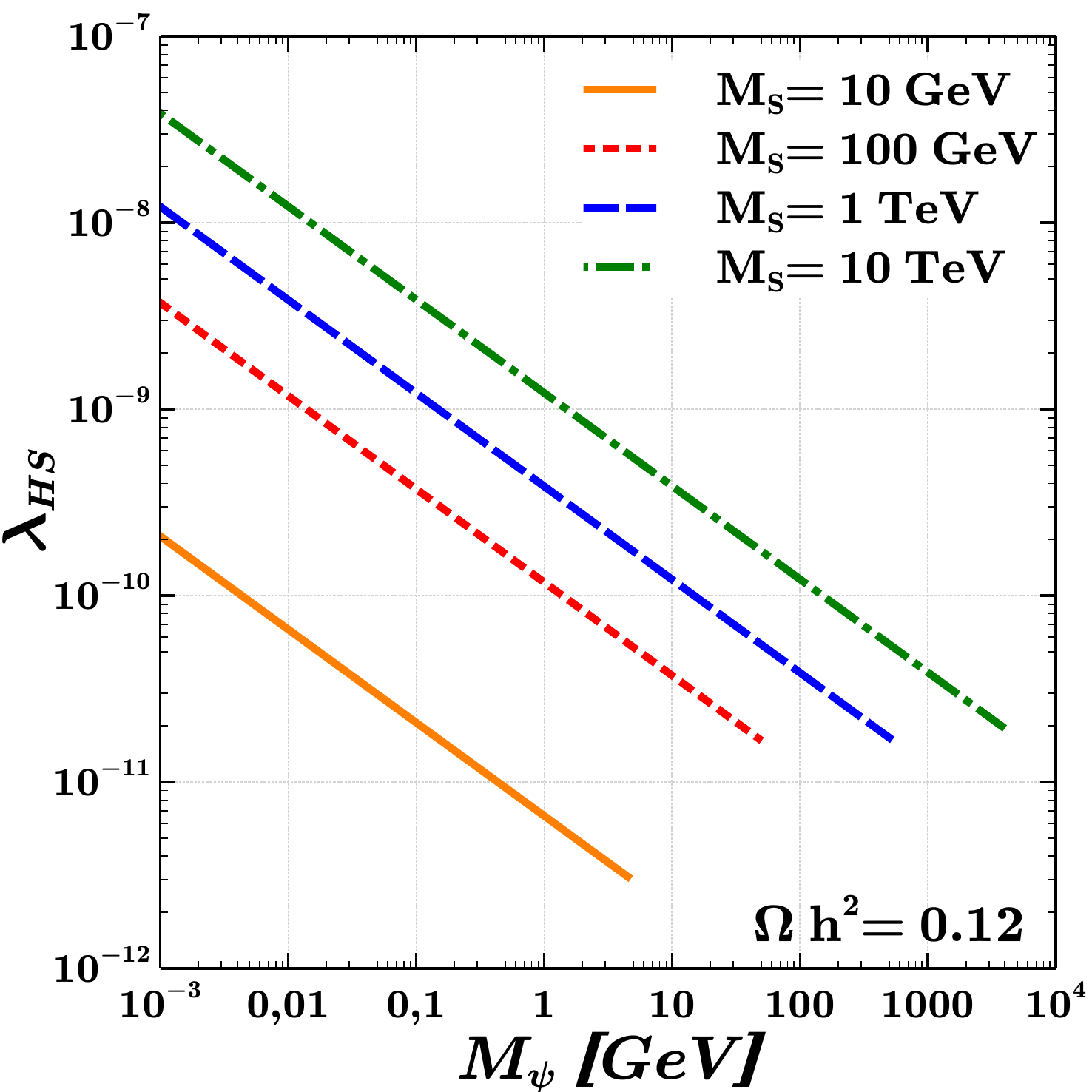}
\caption{The regions consistent with the relic density for case 2 projected onto the planes $(M_S,\lambda_{HS})$ and $(M_\psi,\lambda_{HS})$. The lines differ in length due to the condition $M_S>2M_\psi$. As explained in the text, the values of $y_{s}$ and $y_p$ are irrelevant in this case.  }
\label{fig:pscase2}
\end{figure}

The regions consistent with the dark matter constraint are displayed in figure \ref{fig:pscase2}. The left panel  shows the values of $(M_S, \lambda_{HS})$ that yield $\Omega h^2=0.12$ for different dark matter masses: $1$~GeV (dotted), $10$~GeV (dashed), $100$~GeV (dash-dotted), and $1$~TeV (solid). The first two clearly illustrate the effect of the Higgs resonance discussed in the previous figure. Since $M_S>2M_\psi$, the minimum allowed value of $M_S$ increases with  $M_\psi$--from top to bottom. Notice that, for the range of masses displayed, $\lambda_{HS}$ varies between  $10^{-9}$ and $10^{-12}$. In the right panel the viable regions are instead projected onto the plane $(M_\psi,\lambda_{HS})$ for different values of $M_S$: $10$~GeV (solid), $100$~GeV (dotted), $1$ TeV (dashed), and  $10$~TeV (dash-dotted). As a consequence of the Higgs resonance, the line for $M_S=10$~GeV (below the resonance) is farther apart from the other three lines (above the resonance). In this figure, the maximum value of $M_\psi$ is given by $M_S/2$ whereas its minimum value could be stretched down to about $10$ keV, as in case 1.

\subsection{Case 3}

\begin{figure}[t]
\centering
\includegraphics[scale=0.65]{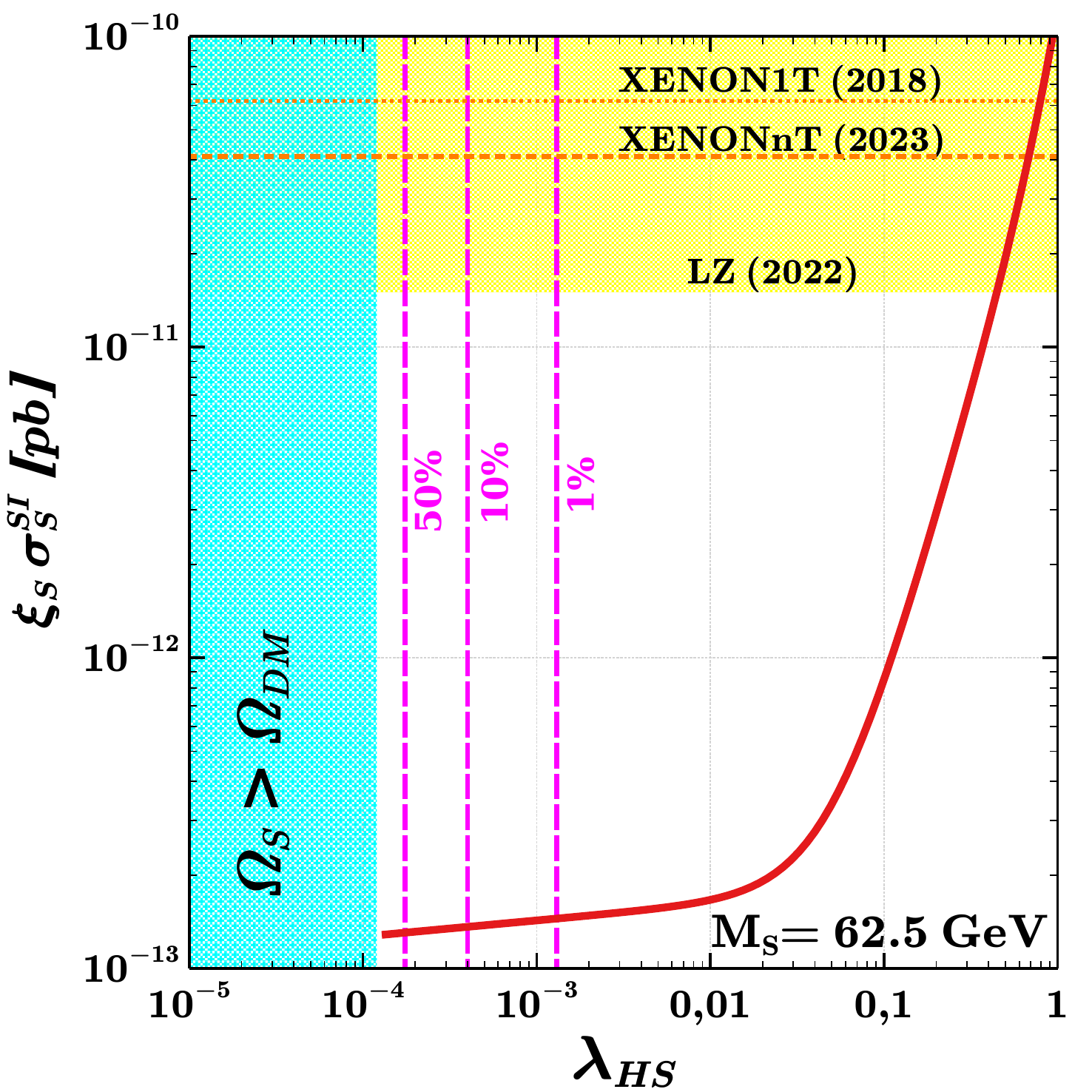}
\caption{The viable parameter space at the Higgs resonance. The solid line shows $\xi_S \sigma^{SI}_S$ as a function of $\lambda_{SH}$ for $M_S=62.5$ GeV. The left part of the figure is excluded because $\Omega_S>\Omega_{DM}$ while the top region is inconsistent with direct detection limits. The vertical magenta lines correspond respectively to $\xi_S=0.5, 0.1, 0.01$.}
\label{fig:psHres}
\end{figure}

In this  case,  $M_S<2M_\psi$, so that both, $S$ and $\psi$, are stable and contribute to the dark matter density --it is a two-component dark matter scenario.  The dark matter constraint then reads
\begin{equation}
    \Omega_{\psi}+\Omega_{S}=\Omega_{\text{DM}}.
\end{equation}
An important  quantity  in these scenarios is the fractional contribution of each  particle to the total dark matter density:
\begin{equation}
\xi_{\psi,S}\equiv \frac{\Omega_{\psi,S}}{\Omega_{\text{DM}}}.    
\end{equation}
Thus, the dark matter constraint can be written as  $\xi_{\psi} +\xi_{S}=1$

Because  $\lambda_{SH}\gtrsim 10^{-6}$ and $y_p,y_s\ll 1$, $\psi$ is a FIMP while  $S$ is a WIMP, yielding a mixed FIMP-WIMP scenario. A consequence of the $S$ relic density being  the result of a freeze-out is that the strong constraints from direct detection experiments apply to it, just as in the singlet scalar model. In fact, the viable regions necessarily lie either at the Higgs resonance, $M_S\sim M_h/2$, or in the multi-TeV range, $M_S\gtrsim 3$ TeV. Here, we will focus on the first possibility, setting $M_S=62.5$ GeV in the following. 

Figure \ref{fig:psHres} shows the viable parameter space (solid red line) at the Higgs resonance in the plane $(\lambda_{HS}, \xi_S\sigma_S^{SI})$ . Notice that  values of $\lambda_{HS}$ close to $1$ are excluded by direct detection limits (yellow region, top), whereas $\lambda_{HS}\lesssim 10^{-4}$ leads to an $\Omega_S$  larger than the observed dark matter density (cyan region, left). The vertical (magenta) dashed lines indicate the values of $\lambda_{HS}$ for which $S$ gives a contribution to the dark matter density of $50\%, 10\%,$ and $1\%$ respectively (from left to right). Accordingly,  both particles ($S$ and $\psi$) contribute significantly to the dark matter density if $\lambda_{HS}$ lies between $10^{-3}$ and $10^{-4}$. 

\begin{figure}[t]
\centering
\includegraphics[scale=0.9]{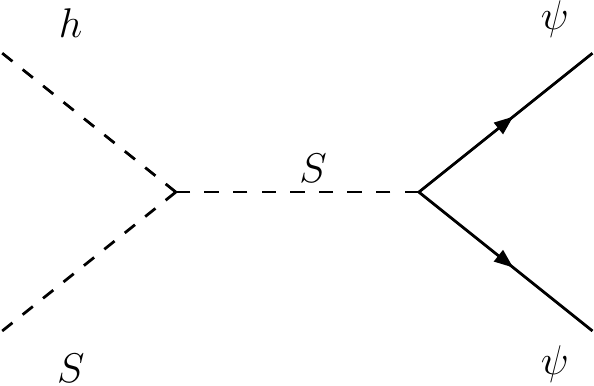}\hspace{0.4cm}
\includegraphics[scale=0.9]{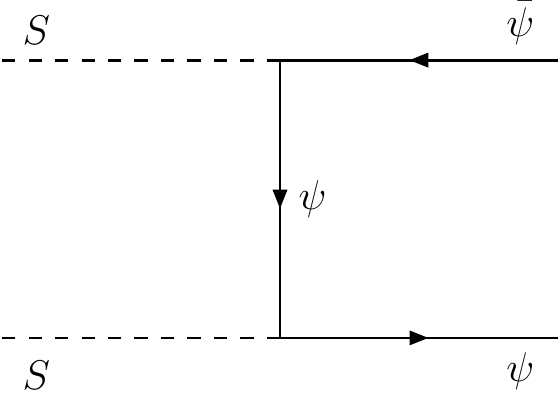}
\caption{Dark matter production processes in case 3: same-charge pair production (left) and conversion (right).}
\label{fig:semi-conv}
\end{figure}

\begin{figure}[t]
\centering
\includegraphics[scale=0.52]{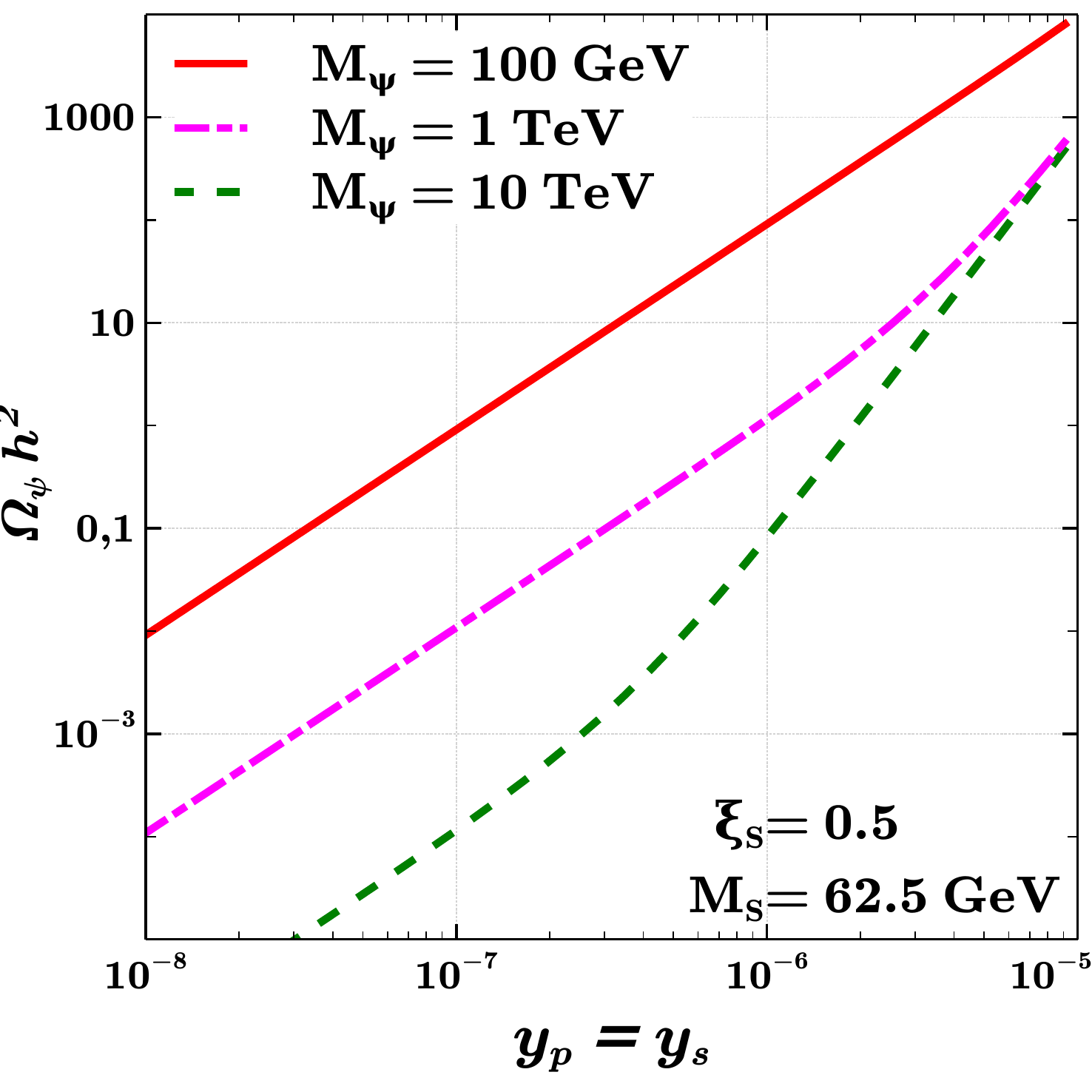}\hspace{4mm}
\includegraphics[scale=0.52]{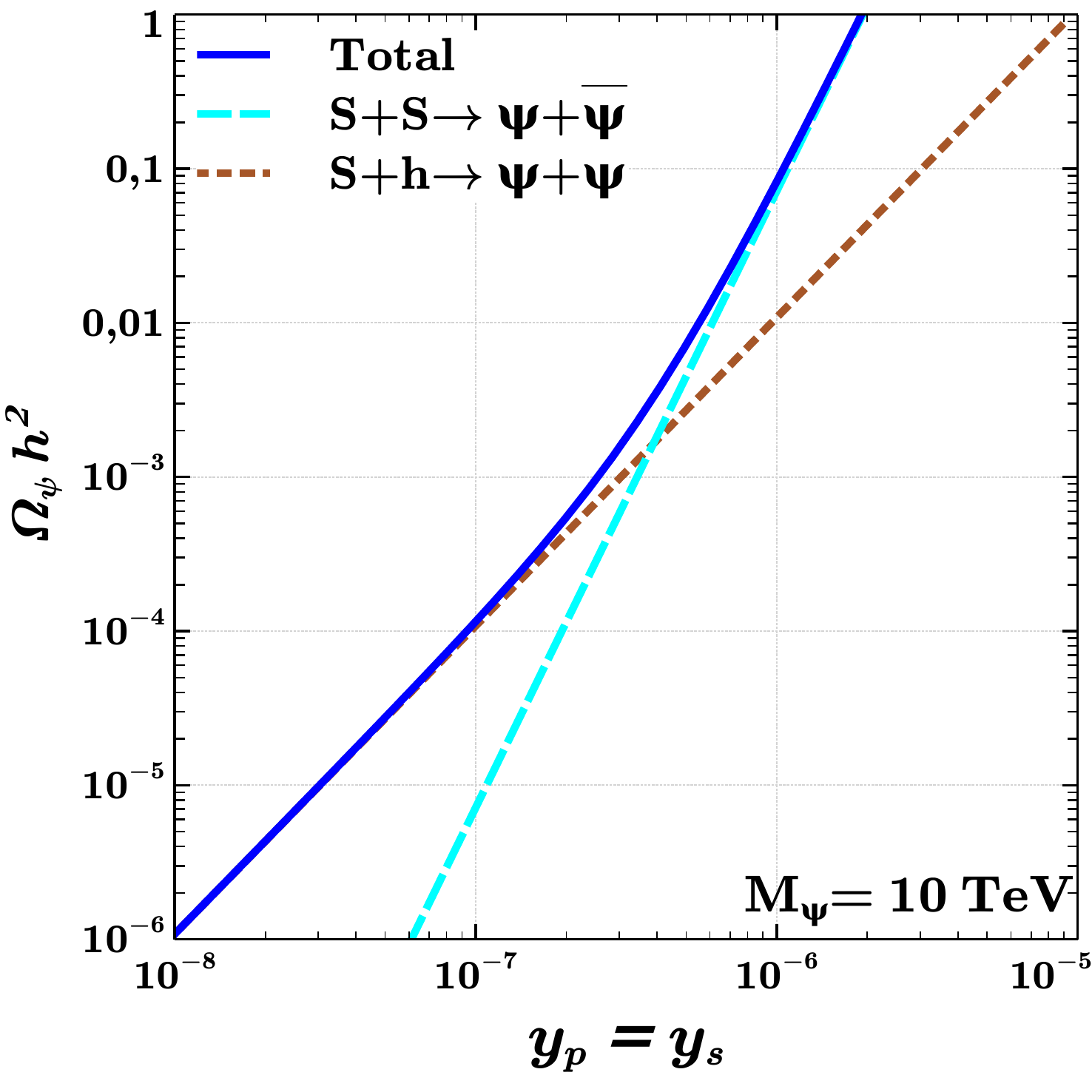}
\caption{Left: The $\psi$ relic density as a function of $y_s=y_p$ for different values of $M_\psi$. Right: The separate contributions of conversion processes (dashed cyan) and same-charge pair production (dotted brown) to the relic density (solid) for $M_\psi=10$ TeV. }
\label{fig:reliccase3}
\end{figure}

Regarding the $\psi$ relic density, which is the result of a freeze-in, two processes contribute to it: $S+S\to \psi+\bar{\psi}$ and $S+h\to \psi+\psi$ (plus its conjugate: $S+h\to \bar\psi+\bar\psi$) --see figure \ref{fig:semi-conv}. In the following, we will label these processes respectively as \emph{conversion} and \emph{same-charge pair production}.  Curiously, $S$ appears in the initial state for both processes. Thus, we have a production mechanism in which FIMPs ($\psi$) are generated from WIMPs ($S$). From their Feynman diagrams, one can see that  $\Omega_\psi\propto y^4$ for conversions while $\Omega_\psi\propto y^2$ for same-charge (SC) pair production.  The left panel of figure \ref{fig:reliccase3} shows the $\psi$ relic density as a function of the Yukawa couplings, which are assumed equal, for different values of $M_\psi$: $100$~GeV (solid line), $1$~TeV (dash-dotted line), and $10$~TeV (dashed line). For $M_\psi=100$ GeV, the relic density increases quadratically with $y_{s,p}$, indicating that it is entirely dominated by $S+h\to \psi+\psi$. A similar result is observed for $M_\psi=1$~TeV, except for large Yukawa couplings ($>10^{-6}$), where the slope becomes steeper due to the contribution from $S+S\to \psi+\bar{\psi}$. For $M_\psi=10$~TeV the transition from SC pair production to conversions is more marked and takes place at smaller values of the Yukawas. To emphasize this point, we separately display, in the right panel, the contributions from SC pair production (dotted line) and conversions (dashed line) to the total relic density (solid line) for $M_\psi=10$~TeV. Notice that SC pair production dominates for $y<10^{-7}$  whereas conversion do so for $y>10^{-6}$.

\begin{figure}[t]
\centering
\includegraphics[scale=0.8]{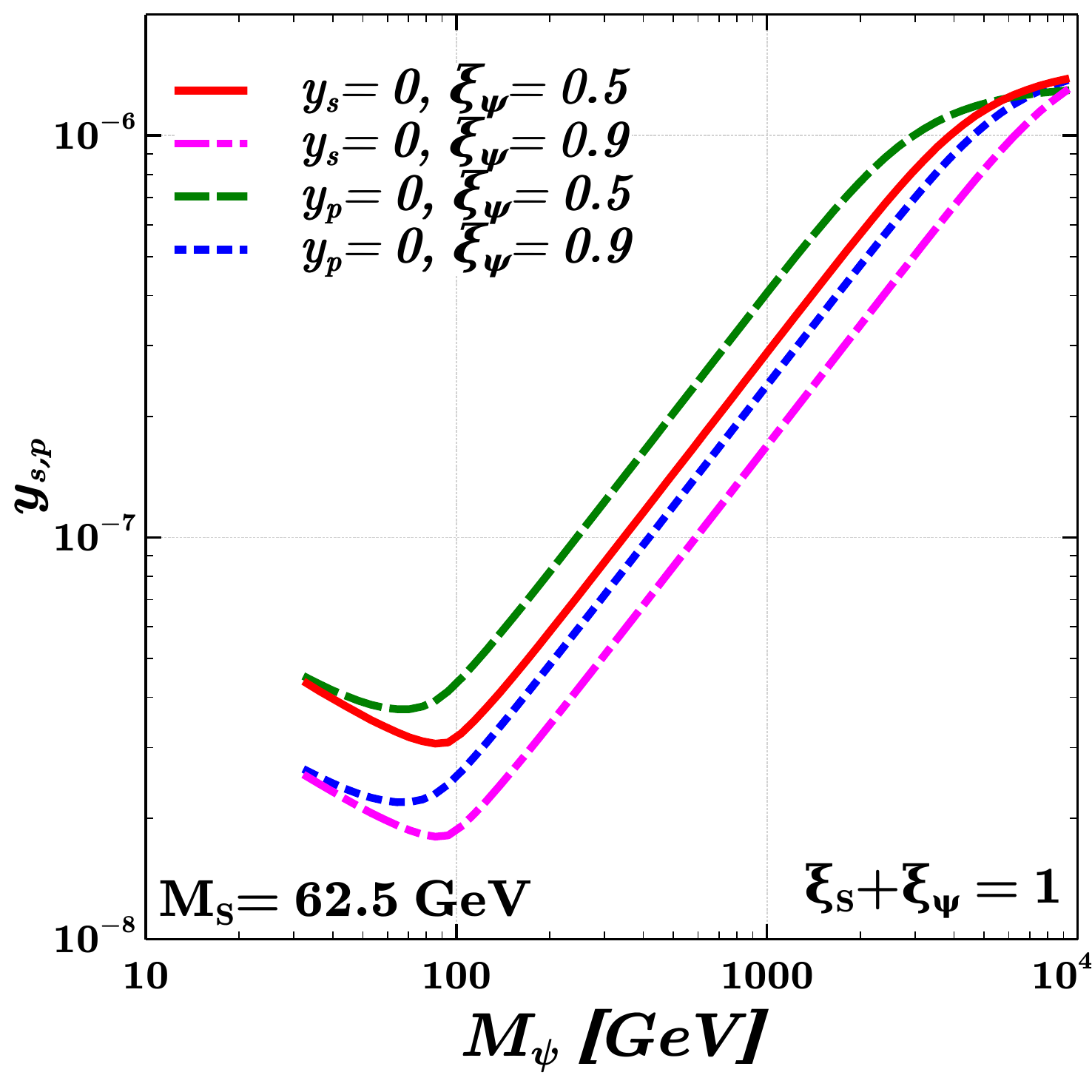}
\caption{The regions consistent with the relic density for case 3 ($M_S=62.5$ GeV). The lines correspond to scalar and pseudoscalar cases for two different values of $\xi_\psi: 0.5, 0.9$. The value of $\lambda_{HS}$ is fixed by the dark matter constraint.}
\label{fig:pscase3}
\end{figure}

Figure \ref{fig:pscase3} illustrates, in the plane $(M_\psi,y_{s,p})$, the  regions consistent with the dark matter constraint for this case. The minimum value of $M_\psi$ is determined by $\frac{M_S}{2}=31.25$~GeV, while its maximum value was set to $10$~TeV. For definiteness, we have considered the scalar ($y_p=0$) and pseudoscalar ($y_s=0$) scenarios for two different values of $\xi_\psi: 0.5, 0.9$. The respective values of $\xi_S$ are therefore $0.5$ and $0.1$, which  set the value of $\lambda_{HS}$ according to figure \ref{fig:psHres}.   In agreement with the previous figures, the Yukawas  tend to increase with $M_\psi$, lying in the interval $(10^{-8},10^{-6})$ for the range  examined. The reason this interval spans couplings larger than in cases 1 and 2 is twofold, depending on the dominant process for dark matter production.  When it is dominated by  SC pair production, there is an additional suppression coming from $\lambda_{HS}\sim 10^{-4}$, and when it is dominated by conversions,  the $y^4$ dependence (rather than $y^2$) of the production cross section permits larger Yukawa couplings.  In any case, note that at a given mass the allowed values of the Yukawas differ by at most a factor 2 or so among the different lines.

\section{Discussion}
\label{sec:discussion}

As we have seen, the $Z_4$ model of fermion FIMP dark matter provides a viable and compelling realization of freeze-in. Let us now show that it is also simpler than other scenarios with  the same particle content --one fermion and one scalar-- that are based on an arbitrary $Z_{N}$ symmetry \cite{Yaguna:2019cvp, Yaguna:2021vhb}. In  the $Z_2$ model, the scalar must be even (a singlet), so it mixes with the Higgs boson, resulting in a model with two additional parameters in the scalar potential~\cite{Klasen:2013ypa}. For the case  $N=3n$,  with $n$  an integer, the scalar potential allows for the inclusion of a trilinear self-interaction term $\mu_3 S^3$,  enlarging the set of free parameters. If, instead,  $N=4n$, the resulting symmetry is equivalent to a $Z_4$. In this scenario, the charges of the fermion $\psi$ and the scalar $S$ are given by $Z_{4n}(\psi)= w_{4n}^{n}=i$ and $Z_{4n}(S)= w_{4n}^{2n}=-1$, respectively. Finally, when $N\neq 4n$, the scalar must be a complex field, introducing an extra degree of freedom in the spectrum. 
We conclude, therefore, that among the $Z_N$ symmetries~\cite{Batell:2010bp,Belanger:2012vp,Belanger:2014bga}, the $Z_4$ we have studied indeed leads to the minimal model. 

Given the small couplings that characterize the viable regions of this model, no dark matter signals are foreseen in any current or planned experiment --a rather generic feature of FIMP dark matter.  Hence, if a signal were actually found, one could immediately  exclude this scenario as the correct explanation of the dark matter. Cosmology offers, nonetheless, an indirect way of probing some regions of parameter space.  In scenarios where the interactions between the SM sector and a hidden sector involving a scalar field $S$ are weakly coupled, such as in our case 2,  the $S$ primordial condensate can act as a second mechanism for dark matter production, and it may also generate isocurvature perturbations~\cite{Nurmi:2015ema, Kainulainen:2016vzv}\footnote{Other possibility may be to consider dark matter production during inflationary reheating~\cite{Bernal:2022wck, Silva-Malpartida:2023yks, Becker:2023tvd}.}.  In addition, it would be  interesting to explore the conditions under which the scalar field $S$ could potentially drive cosmic inflation, reheat the Universe, and produce the observed abundance of fermion FIMP dark matter. Such considerations might involve the introduction of sizeable non-minimal curvature couplings or other mechanisms to facilitate the desired cosmological evolution~\cite{Tenkanen:2016twd,Almeida:2018oid}. Studying these aspects may provide valuable insights into the model's broader cosmological implications, but it is unlikely it will alter the conclusions reached in the previous section. In any case, such analysis  lies beyond the scope of this work.

A simple and motivated extension of the model we discussed can be attained by promoting the $Z_4$ symmetry to a $U(1)'$ gauge symmetry. In that case, the scalar field would become  complex and the Lagrangian would be given by:
\begin{align}\label{eq:U1lag}
 \mathcal{L}&=-\frac{1}{4}F'_{\mu\nu}F'^{\mu\nu}+\overline{\psi}(i\gamma^\mu\partial_\mu-g'q_\psi\gamma^\mu A'_\mu -M_\psi)\psi + |(\partial_\mu +ig'q_S A'_\mu)S|^2-\mu_{S}^2|S|^2 - \lambda_{S}|S|^4 \nonumber\\
 & -\lambda_{S H}|H|^2|S|^2+\frac{1}{2}\left[y_s\overline{\psi^c}\psi S + y_{p}\overline{\psi^c}\gamma_5\psi S + \rm{h.c.}\right]+\epsilon F_{Y\mu\nu}F'^{\mu\nu}, 
 \end{align}
where $q_S=-2q_\psi$. In order to achieve fermion FIMP dark matter, both the $U(1)'$ gauge coupling $g'$ and the kinetic mixing parameter $\epsilon$ must satisfy $g',\epsilon\ll y_{s,p}$, which enables the scenarios discussed in the previous section (cases 1 to 3). However, it is important to note that if $g'\sim y_{s,p}$ and $\epsilon$ is sizeable, the $A'$ gauge boson may also contribute significantly to the fermion relic abundance through the decay $A'\to \bar{\psi}+\psi$. In such cases, additional considerations and constraints are necessary to accurately determine the contributions of both the scalar $S$ and the $A'$ to the overall dark matter abundance.

\section{Conclusions}\label{sec:conclusions}
We studied  a new realization of the freeze-in mechanism for fermion dark matter. The model is quite simple as it extends the SM with just a $Z_4$ symmetry and two new  fields, $S$ and $\psi$,  charged under it. In fact, only 5  parameters --$M_S$, $M_\psi$, $y_p$, $y_s$ and $\lambda_{HS}$-- determine its phenomenology.  Within this scenario, which had not been studied before in the context of FIMP dark matter, we identified three independent instances of freeze-in, each featuring its own dark matter production mechanism: $i)$ the decay $S\to \psi+\psi$ while $S$ is in equilibrium; $ii)$ pair-production of $S$ followed by its decay while out of equilibrium; $iii)$ FIMP from WIMP via the processes $S+h\to \psi+\psi$ and $S+S\to \psi+\bar\psi$. The first two cases feature FIMP dark matter whereas the third one is a two-component mixed FIMP-WIMP scenario. For each of these cases  we studied the relic density as a function of the parameters of the model and determined the regions of parameter space that are consistent with the observed dark matter abundance.  Our results demonstrate that this scenario offers a phenomenologically rich setup that is viable  over a wide range of masses and couplings. In addition, it furnishes a minimal model of freeze-in for fermion FIMP dark matter.   

\section*{Acknowledgments}
OZ and CY received funding from the Patrimonio Autónomo - Fondo Nacional de Financiamiento para la Ciencia, la Tecnología y la Innovación Francisco José de Caldas (MinCiencias - Colombia) grant 82315-2021-1080.  OZ has received further funding from MinCiencias through the Grant 80740-492-2021, and has been partially supported by Sostenibilidad-UdeA and the UdeA/CODI Grant 2020-33177.  OZ is grateful to the Physics School of Universidad Pedagógica y Tecnológica de Tunja for the warm hospitality extended during the completion of this work.

\bibliography{references}

\begin{thebibliography}{46}%
\makeatletter
\providecommand \@ifxundefined [1]{%
 \@ifx{#1\undefined}
}%
\providecommand \@ifnum [1]{%
 \ifnum #1\expandafter \@firstoftwo
 \else \expandafter \@secondoftwo
 \fi
}%
\providecommand \@ifx [1]{%
 \ifx #1\expandafter \@firstoftwo
 \else \expandafter \@secondoftwo
 \fi
}%
\providecommand \natexlab [1]{#1}%
\providecommand \enquote  [1]{``#1''}%
\providecommand \bibnamefont  [1]{#1}%
\providecommand \bibfnamefont [1]{#1}%
\providecommand \citenamefont [1]{#1}%
\providecommand \href@noop [0]{\@secondoftwo}%
\providecommand \href [0]{\begingroup \@sanitize@url \@href}%
\providecommand \@href[1]{\@@startlink{#1}\@@href}%
\providecommand \@@href[1]{\endgroup#1\@@endlink}%
\providecommand \@sanitize@url [0]{\catcode `\\12\catcode `\$12\catcode
  `\&12\catcode `\#12\catcode `\^12\catcode `\_12\catcode `\%12\relax}%
\providecommand \@@startlink[1]{}%
\providecommand \@@endlink[0]{}%
\providecommand \url  [0]{\begingroup\@sanitize@url \@url }%
\providecommand \@url [1]{\endgroup\@href {#1}{\urlprefix }}%
\providecommand \urlprefix  [0]{URL }%
\providecommand \Eprint [0]{\href }%
\providecommand \doibase [0]{http://dx.doi.org/}%
\providecommand \selectlanguage [0]{\@gobble}%
\providecommand \bibinfo  [0]{\@secondoftwo}%
\providecommand \bibfield  [0]{\@secondoftwo}%
\providecommand \translation [1]{[#1]}%
\providecommand \BibitemOpen [0]{}%
\providecommand \bibitemStop [0]{}%
\providecommand \bibitemNoStop [0]{.\EOS\space}%
\providecommand \EOS [0]{\spacefactor3000\relax}%
\providecommand \BibitemShut  [1]{\csname bibitem#1\endcsname}%
\let\auto@bib@innerbib\@empty
\bibitem [{\citenamefont {Hinshaw}\ \emph {et~al.}(2013)\citenamefont {Hinshaw}
  \emph {et~al.}}]{WMAP:2012nax}%
  \BibitemOpen
  \bibfield  {author} {\bibinfo {author} {\bibfnamefont {G.}~\bibnamefont
  {Hinshaw}} \emph {et~al.} (\bibinfo {collaboration} {WMAP}),\ }\href
  {\doibase 10.1088/0067-0049/208/2/19} {\bibfield  {journal} {\bibinfo
  {journal} {Astrophys. J. Suppl.}\ }\textbf {\bibinfo {volume} {208}},\
  \bibinfo {pages} {19} (\bibinfo {year} {2013})},\ \Eprint
  {http://arxiv.org/abs/1212.5226} {arXiv:1212.5226 [astro-ph.CO]} \BibitemShut
  {NoStop}%
\bibitem [{\citenamefont {Aghanim}\ \emph {et~al.}(2020)\citenamefont {Aghanim}
  \emph {et~al.}}]{Aghanim:2018eyx}%
  \BibitemOpen
  \bibfield  {author} {\bibinfo {author} {\bibfnamefont {N.}~\bibnamefont
  {Aghanim}} \emph {et~al.} (\bibinfo {collaboration} {Planck}),\ }\href
  {\doibase 10.1051/0004-6361/201833910} {\bibfield  {journal} {\bibinfo
  {journal} {Astron. Astrophys.}\ }\textbf {\bibinfo {volume} {641}},\ \bibinfo
  {pages} {A6} (\bibinfo {year} {2020})},\ \Eprint
  {http://arxiv.org/abs/1807.06209} {arXiv:1807.06209 [astro-ph.CO]}
  \BibitemShut {NoStop}%
\bibitem [{\citenamefont {Bernal}\ \emph {et~al.}(2017)\citenamefont {Bernal},
  \citenamefont {Heikinheimo}, \citenamefont {Tenkanen}, \citenamefont
  {Tuominen},\ and\ \citenamefont {Vaskonen}}]{Bernal:2017kxu}%
  \BibitemOpen
  \bibfield  {author} {\bibinfo {author} {\bibfnamefont {N.}~\bibnamefont
  {Bernal}}, \bibinfo {author} {\bibfnamefont {M.}~\bibnamefont {Heikinheimo}},
  \bibinfo {author} {\bibfnamefont {T.}~\bibnamefont {Tenkanen}}, \bibinfo
  {author} {\bibfnamefont {K.}~\bibnamefont {Tuominen}}, \ and\ \bibinfo
  {author} {\bibfnamefont {V.}~\bibnamefont {Vaskonen}},\ }\href {\doibase
  10.1142/S0217751X1730023X} {\bibfield  {journal} {\bibinfo  {journal} {Int.
  J. Mod. Phys.}\ }\textbf {\bibinfo {volume} {A32}},\ \bibinfo {pages}
  {1730023} (\bibinfo {year} {2017})},\ \Eprint
  {http://arxiv.org/abs/1706.07442} {arXiv:1706.07442 [hep-ph]} \BibitemShut
  {NoStop}%
\bibitem [{\citenamefont {Agrawal}\ \emph {et~al.}(2021)\citenamefont {Agrawal}
  \emph {et~al.}}]{Agrawal:2021dbo}%
  \BibitemOpen
  \bibfield  {author} {\bibinfo {author} {\bibfnamefont {P.}~\bibnamefont
  {Agrawal}} \emph {et~al.},\ }\href {\doibase 10.1140/epjc/s10052-021-09703-7}
  {\bibfield  {journal} {\bibinfo  {journal} {Eur. Phys. J. C}\ }\textbf
  {\bibinfo {volume} {81}},\ \bibinfo {pages} {1015} (\bibinfo {year}
  {2021})},\ \Eprint {http://arxiv.org/abs/2102.12143} {arXiv:2102.12143
  [hep-ph]} \BibitemShut {NoStop}%
\bibitem [{\citenamefont {Hall}\ \emph {et~al.}(2010)\citenamefont {Hall},
  \citenamefont {Jedamzik}, \citenamefont {March-Russell},\ and\ \citenamefont
  {West}}]{Hall:2009bx}%
  \BibitemOpen
  \bibfield  {author} {\bibinfo {author} {\bibfnamefont {L.~J.}\ \bibnamefont
  {Hall}}, \bibinfo {author} {\bibfnamefont {K.}~\bibnamefont {Jedamzik}},
  \bibinfo {author} {\bibfnamefont {J.}~\bibnamefont {March-Russell}}, \ and\
  \bibinfo {author} {\bibfnamefont {S.~M.}\ \bibnamefont {West}},\ }\href
  {\doibase 10.1007/JHEP03(2010)080} {\bibfield  {journal} {\bibinfo  {journal}
  {JHEP}\ }\textbf {\bibinfo {volume} {03}},\ \bibinfo {pages} {080} (\bibinfo
  {year} {2010})},\ \Eprint {http://arxiv.org/abs/0911.1120} {arXiv:0911.1120
  [hep-ph]} \BibitemShut {NoStop}%
\bibitem [{\citenamefont {Steigman}\ \emph {et~al.}(2012)\citenamefont
  {Steigman}, \citenamefont {Dasgupta},\ and\ \citenamefont
  {Beacom}}]{Steigman:2012nb}%
  \BibitemOpen
  \bibfield  {author} {\bibinfo {author} {\bibfnamefont {G.}~\bibnamefont
  {Steigman}}, \bibinfo {author} {\bibfnamefont {B.}~\bibnamefont {Dasgupta}},
  \ and\ \bibinfo {author} {\bibfnamefont {J.~F.}\ \bibnamefont {Beacom}},\
  }\href {\doibase 10.1103/PhysRevD.86.023506} {\bibfield  {journal} {\bibinfo
  {journal} {Phys. Rev.}\ }\textbf {\bibinfo {volume} {D86}},\ \bibinfo {pages}
  {023506} (\bibinfo {year} {2012})},\ \Eprint {http://arxiv.org/abs/1204.3622}
  {arXiv:1204.3622 [hep-ph]} \BibitemShut {NoStop}%
\bibitem [{\citenamefont {Bélanger}\ \emph {et~al.}(2018)\citenamefont
  {Bélanger}, \citenamefont {Boudjema}, \citenamefont {Goudelis},
  \citenamefont {Pukhov},\ and\ \citenamefont {Zaldivar}}]{Belanger:2018ccd}%
  \BibitemOpen
  \bibfield  {author} {\bibinfo {author} {\bibfnamefont {G.}~\bibnamefont
  {Bélanger}}, \bibinfo {author} {\bibfnamefont {F.}~\bibnamefont {Boudjema}},
  \bibinfo {author} {\bibfnamefont {A.}~\bibnamefont {Goudelis}}, \bibinfo
  {author} {\bibfnamefont {A.}~\bibnamefont {Pukhov}}, \ and\ \bibinfo {author}
  {\bibfnamefont {B.}~\bibnamefont {Zaldivar}},\ }\href {\doibase
  10.1016/j.cpc.2018.04.027} {\bibfield  {journal} {\bibinfo  {journal}
  {Comput. Phys. Commun.}\ }\textbf {\bibinfo {volume} {231}},\ \bibinfo
  {pages} {173} (\bibinfo {year} {2018})},\ \Eprint
  {http://arxiv.org/abs/1801.03509} {arXiv:1801.03509 [hep-ph]} \BibitemShut
  {NoStop}%
\bibitem [{\citenamefont {Bringmann}\ \emph {et~al.}(2022)\citenamefont
  {Bringmann}, \citenamefont {Heeba}, \citenamefont {Kahlhoefer},\ and\
  \citenamefont {Vangsnes}}]{Bringmann:2021sth}%
  \BibitemOpen
  \bibfield  {author} {\bibinfo {author} {\bibfnamefont {T.}~\bibnamefont
  {Bringmann}}, \bibinfo {author} {\bibfnamefont {S.}~\bibnamefont {Heeba}},
  \bibinfo {author} {\bibfnamefont {F.}~\bibnamefont {Kahlhoefer}}, \ and\
  \bibinfo {author} {\bibfnamefont {K.}~\bibnamefont {Vangsnes}},\ }\href
  {\doibase 10.1007/JHEP02(2022)110} {\bibfield  {journal} {\bibinfo  {journal}
  {JHEP}\ }\textbf {\bibinfo {volume} {02}},\ \bibinfo {pages} {110} (\bibinfo
  {year} {2022})},\ \Eprint {http://arxiv.org/abs/2111.14871} {arXiv:2111.14871
  [hep-ph]} \BibitemShut {NoStop}%
\bibitem [{\citenamefont {Bertone}\ and\ \citenamefont
  {Tait}(2018)}]{Bertone:2018krk}%
  \BibitemOpen
  \bibfield  {author} {\bibinfo {author} {\bibfnamefont {G.}~\bibnamefont
  {Bertone}}\ and\ \bibinfo {author} {\bibfnamefont {T.}~\bibnamefont {Tait},
  \bibfnamefont {M.~P.}},\ }\href {\doibase 10.1038/s41586-018-0542-z}
  {\bibfield  {journal} {\bibinfo  {journal} {Nature}\ }\textbf {\bibinfo
  {volume} {562}},\ \bibinfo {pages} {51} (\bibinfo {year} {2018})},\ \Eprint
  {http://arxiv.org/abs/1810.01668} {arXiv:1810.01668 [astro-ph.CO]}
  \BibitemShut {NoStop}%
\bibitem [{\citenamefont {Yaguna}(2011)}]{Yaguna:2011qn}%
  \BibitemOpen
  \bibfield  {author} {\bibinfo {author} {\bibfnamefont {C.~E.}\ \bibnamefont
  {Yaguna}},\ }\href {\doibase 10.1007/JHEP08(2011)060} {\bibfield  {journal}
  {\bibinfo  {journal} {JHEP}\ }\textbf {\bibinfo {volume} {08}},\ \bibinfo
  {pages} {060} (\bibinfo {year} {2011})},\ \Eprint
  {http://arxiv.org/abs/1105.1654} {arXiv:1105.1654 [hep-ph]} \BibitemShut
  {NoStop}%
\bibitem [{\citenamefont {Silveira}\ and\ \citenamefont
  {Zee}(1985)}]{Silveira:1985rk}%
  \BibitemOpen
  \bibfield  {author} {\bibinfo {author} {\bibfnamefont {V.}~\bibnamefont
  {Silveira}}\ and\ \bibinfo {author} {\bibfnamefont {A.}~\bibnamefont {Zee}},\
  }\href {\doibase 10.1016/0370-2693(85)90624-0} {\bibfield  {journal}
  {\bibinfo  {journal} {Phys. Lett. B}\ }\textbf {\bibinfo {volume} {161}},\
  \bibinfo {pages} {136} (\bibinfo {year} {1985})}\BibitemShut {NoStop}%
\bibitem [{\citenamefont {McDonald}(1994)}]{McDonald:1993ex}%
  \BibitemOpen
  \bibfield  {author} {\bibinfo {author} {\bibfnamefont {J.}~\bibnamefont
  {McDonald}},\ }\href {\doibase 10.1103/PhysRevD.50.3637} {\bibfield
  {journal} {\bibinfo  {journal} {Phys. Rev.}\ }\textbf {\bibinfo {volume}
  {D50}},\ \bibinfo {pages} {3637} (\bibinfo {year} {1994})},\ \Eprint
  {http://arxiv.org/abs/hep-ph/0702143} {arXiv:hep-ph/0702143 [HEP-PH]}
  \BibitemShut {NoStop}%
\bibitem [{\citenamefont {Burgess}\ \emph {et~al.}(2001)\citenamefont
  {Burgess}, \citenamefont {Pospelov},\ and\ \citenamefont {ter
  Veldhuis}}]{Burgess:2000yq}%
  \BibitemOpen
  \bibfield  {author} {\bibinfo {author} {\bibfnamefont {C.}~\bibnamefont
  {Burgess}}, \bibinfo {author} {\bibfnamefont {M.}~\bibnamefont {Pospelov}}, \
  and\ \bibinfo {author} {\bibfnamefont {T.}~\bibnamefont {ter Veldhuis}},\
  }\href {\doibase 10.1016/S0550-3213(01)00513-2} {\bibfield  {journal}
  {\bibinfo  {journal} {Nucl. Phys. B}\ }\textbf {\bibinfo {volume} {619}},\
  \bibinfo {pages} {709} (\bibinfo {year} {2001})},\ \Eprint
  {http://arxiv.org/abs/hep-ph/0011335} {arXiv:hep-ph/0011335} \BibitemShut
  {NoStop}%
\bibitem [{\citenamefont {Kim}\ and\ \citenamefont {Lee}(2007)}]{Kim:2006af}%
  \BibitemOpen
  \bibfield  {author} {\bibinfo {author} {\bibfnamefont {Y.~G.}\ \bibnamefont
  {Kim}}\ and\ \bibinfo {author} {\bibfnamefont {K.~Y.}\ \bibnamefont {Lee}},\
  }\href {\doibase 10.1103/PhysRevD.75.115012} {\bibfield  {journal} {\bibinfo
  {journal} {Phys. Rev. D}\ }\textbf {\bibinfo {volume} {75}},\ \bibinfo
  {pages} {115012} (\bibinfo {year} {2007})},\ \Eprint
  {http://arxiv.org/abs/hep-ph/0611069} {arXiv:hep-ph/0611069} \BibitemShut
  {NoStop}%
\bibitem [{\citenamefont {Kim}\ \emph {et~al.}(2008)\citenamefont {Kim},
  \citenamefont {Lee},\ and\ \citenamefont {Shin}}]{Kim:2008pp}%
  \BibitemOpen
  \bibfield  {author} {\bibinfo {author} {\bibfnamefont {Y.~G.}\ \bibnamefont
  {Kim}}, \bibinfo {author} {\bibfnamefont {K.~Y.}\ \bibnamefont {Lee}}, \ and\
  \bibinfo {author} {\bibfnamefont {S.}~\bibnamefont {Shin}},\ }\href {\doibase
  10.1088/1126-6708/2008/05/100} {\bibfield  {journal} {\bibinfo  {journal}
  {JHEP}\ }\textbf {\bibinfo {volume} {05}},\ \bibinfo {pages} {100} (\bibinfo
  {year} {2008})},\ \Eprint {http://arxiv.org/abs/0803.2932} {arXiv:0803.2932
  [hep-ph]} \BibitemShut {NoStop}%
\bibitem [{\citenamefont {Lopez-Honorez}\ \emph {et~al.}(2012)\citenamefont
  {Lopez-Honorez}, \citenamefont {Schwetz},\ and\ \citenamefont
  {Zupan}}]{Lopez-Honorez:2012tov}%
  \BibitemOpen
  \bibfield  {author} {\bibinfo {author} {\bibfnamefont {L.}~\bibnamefont
  {Lopez-Honorez}}, \bibinfo {author} {\bibfnamefont {T.}~\bibnamefont
  {Schwetz}}, \ and\ \bibinfo {author} {\bibfnamefont {J.}~\bibnamefont
  {Zupan}},\ }\href {\doibase 10.1016/j.physletb.2012.07.017} {\bibfield
  {journal} {\bibinfo  {journal} {Phys. Lett. B}\ }\textbf {\bibinfo {volume}
  {716}},\ \bibinfo {pages} {179} (\bibinfo {year} {2012})},\ \Eprint
  {http://arxiv.org/abs/1203.2064} {arXiv:1203.2064 [hep-ph]} \BibitemShut
  {NoStop}%
\bibitem [{\citenamefont {Esch}\ \emph {et~al.}(2013)\citenamefont {Esch},
  \citenamefont {Klasen},\ and\ \citenamefont {Yaguna}}]{Esch:2013rta}%
  \BibitemOpen
  \bibfield  {author} {\bibinfo {author} {\bibfnamefont {S.}~\bibnamefont
  {Esch}}, \bibinfo {author} {\bibfnamefont {M.}~\bibnamefont {Klasen}}, \ and\
  \bibinfo {author} {\bibfnamefont {C.~E.}\ \bibnamefont {Yaguna}},\ }\href
  {\doibase 10.1103/PhysRevD.88.075017} {\bibfield  {journal} {\bibinfo
  {journal} {Phys. Rev. D}\ }\textbf {\bibinfo {volume} {88}},\ \bibinfo
  {pages} {075017} (\bibinfo {year} {2013})},\ \Eprint
  {http://arxiv.org/abs/1308.0951} {arXiv:1308.0951 [hep-ph]} \BibitemShut
  {NoStop}%
\bibitem [{\citenamefont {Klasen}\ and\ \citenamefont
  {Yaguna}(2013)}]{Klasen:2013ypa}%
  \BibitemOpen
  \bibfield  {author} {\bibinfo {author} {\bibfnamefont {M.}~\bibnamefont
  {Klasen}}\ and\ \bibinfo {author} {\bibfnamefont {C.~E.}\ \bibnamefont
  {Yaguna}},\ }\href {\doibase 10.1088/1475-7516/2013/11/039} {\bibfield
  {journal} {\bibinfo  {journal} {JCAP}\ }\textbf {\bibinfo {volume} {1311}},\
  \bibinfo {pages} {039} (\bibinfo {year} {2013})},\ \Eprint
  {http://arxiv.org/abs/1309.2777} {arXiv:1309.2777 [hep-ph]} \BibitemShut
  {NoStop}%
\bibitem [{\citenamefont {Cai}\ and\ \citenamefont
  {Spray}(2016)}]{Cai:2015zza}%
  \BibitemOpen
  \bibfield  {author} {\bibinfo {author} {\bibfnamefont {Y.}~\bibnamefont
  {Cai}}\ and\ \bibinfo {author} {\bibfnamefont {A.~P.}\ \bibnamefont
  {Spray}},\ }\href {\doibase 10.1007/JHEP01(2016)087} {\bibfield  {journal}
  {\bibinfo  {journal} {JHEP}\ }\textbf {\bibinfo {volume} {01}},\ \bibinfo
  {pages} {087} (\bibinfo {year} {2016})},\ \Eprint
  {http://arxiv.org/abs/1509.08481} {arXiv:1509.08481 [hep-ph]} \BibitemShut
  {NoStop}%
\bibitem [{\citenamefont {Yaguna}\ and\ \citenamefont
  {Zapata}(2022)}]{Yaguna:2021rds}%
  \BibitemOpen
  \bibfield  {author} {\bibinfo {author} {\bibfnamefont {C.~E.}\ \bibnamefont
  {Yaguna}}\ and\ \bibinfo {author} {\bibfnamefont {O.}~\bibnamefont
  {Zapata}},\ }\href {\doibase 10.1103/PhysRevD.105.095026} {\bibfield
  {journal} {\bibinfo  {journal} {Phys. Rev. D}\ }\textbf {\bibinfo {volume}
  {105}},\ \bibinfo {pages} {095026} (\bibinfo {year} {2022})},\ \Eprint
  {http://arxiv.org/abs/2112.07020} {arXiv:2112.07020 [hep-ph]} \BibitemShut
  {NoStop}%
\bibitem [{\citenamefont {Barman}\ \emph {et~al.}(2020)\citenamefont {Barman},
  \citenamefont {Dutta~Banik},\ and\ \citenamefont {Paul}}]{Barman:2020jrf}%
  \BibitemOpen
  \bibfield  {author} {\bibinfo {author} {\bibfnamefont {B.}~\bibnamefont
  {Barman}}, \bibinfo {author} {\bibfnamefont {A.}~\bibnamefont {Dutta~Banik}},
  \ and\ \bibinfo {author} {\bibfnamefont {A.}~\bibnamefont {Paul}},\
  }\href@noop {} {\  (\bibinfo {year} {2020})},\ \Eprint
  {http://arxiv.org/abs/2012.11969} {arXiv:2012.11969 [astro-ph.CO]}
  \BibitemShut {NoStop}%
\bibitem [{\citenamefont {Finkbeiner}\ and\ \citenamefont
  {Weiner}(2007)}]{Finkbeiner:2007kk}%
  \BibitemOpen
  \bibfield  {author} {\bibinfo {author} {\bibfnamefont {D.~P.}\ \bibnamefont
  {Finkbeiner}}\ and\ \bibinfo {author} {\bibfnamefont {N.}~\bibnamefont
  {Weiner}},\ }\href {\doibase 10.1103/PhysRevD.76.083519} {\bibfield
  {journal} {\bibinfo  {journal} {Phys. Rev. D}\ }\textbf {\bibinfo {volume}
  {76}},\ \bibinfo {pages} {083519} (\bibinfo {year} {2007})},\ \Eprint
  {http://arxiv.org/abs/astro-ph/0702587} {arXiv:astro-ph/0702587} \BibitemShut
  {NoStop}%
\bibitem [{\citenamefont {Bell}\ \emph {et~al.}(2011)\citenamefont {Bell},
  \citenamefont {Galea},\ and\ \citenamefont {Volkas}}]{Bell:2010qt}%
  \BibitemOpen
  \bibfield  {author} {\bibinfo {author} {\bibfnamefont {N.~F.}\ \bibnamefont
  {Bell}}, \bibinfo {author} {\bibfnamefont {A.~J.}\ \bibnamefont {Galea}}, \
  and\ \bibinfo {author} {\bibfnamefont {R.~R.}\ \bibnamefont {Volkas}},\
  }\href {\doibase 10.1103/PhysRevD.83.063504} {\bibfield  {journal} {\bibinfo
  {journal} {Phys. Rev. D}\ }\textbf {\bibinfo {volume} {83}},\ \bibinfo
  {pages} {063504} (\bibinfo {year} {2011})},\ \Eprint
  {http://arxiv.org/abs/1012.0067} {arXiv:1012.0067 [hep-ph]} \BibitemShut
  {NoStop}%
\bibitem [{\citenamefont {Bhattacharya}\ \emph {et~al.}(2018)\citenamefont
  {Bhattacharya}, \citenamefont {de~Medeiros~Varzielas}, \citenamefont
  {Karmakar}, \citenamefont {King},\ and\ \citenamefont
  {Sil}}]{Bhattacharya:2018ljs}%
  \BibitemOpen
  \bibfield  {author} {\bibinfo {author} {\bibfnamefont {S.}~\bibnamefont
  {Bhattacharya}}, \bibinfo {author} {\bibfnamefont {I.}~\bibnamefont
  {de~Medeiros~Varzielas}}, \bibinfo {author} {\bibfnamefont {B.}~\bibnamefont
  {Karmakar}}, \bibinfo {author} {\bibfnamefont {S.~F.}\ \bibnamefont {King}},
  \ and\ \bibinfo {author} {\bibfnamefont {A.}~\bibnamefont {Sil}},\ }\href
  {\doibase 10.1007/JHEP12(2018)007} {\bibfield  {journal} {\bibinfo  {journal}
  {JHEP}\ }\textbf {\bibinfo {volume} {12}},\ \bibinfo {pages} {007} (\bibinfo
  {year} {2018})},\ \Eprint {http://arxiv.org/abs/1806.00490} {arXiv:1806.00490
  [hep-ph]} \BibitemShut {NoStop}%
\bibitem [{\citenamefont {Belanger}\ \emph {et~al.}(2022)\citenamefont
  {Belanger}, \citenamefont {Mjallal},\ and\ \citenamefont
  {Pukhov}}]{Belanger:2022qxt}%
  \BibitemOpen
  \bibfield  {author} {\bibinfo {author} {\bibfnamefont {G.}~\bibnamefont
  {Belanger}}, \bibinfo {author} {\bibfnamefont {A.}~\bibnamefont {Mjallal}}, \
  and\ \bibinfo {author} {\bibfnamefont {A.}~\bibnamefont {Pukhov}},\ }\href
  {\doibase 10.1103/PhysRevD.106.095019} {\bibfield  {journal} {\bibinfo
  {journal} {Phys. Rev. D}\ }\textbf {\bibinfo {volume} {106}},\ \bibinfo
  {pages} {095019} (\bibinfo {year} {2022})},\ \Eprint
  {http://arxiv.org/abs/2205.04101} {arXiv:2205.04101 [hep-ph]} \BibitemShut
  {NoStop}%
\bibitem [{\citenamefont {Dutta~Banik}\ \emph {et~al.}(2017)\citenamefont
  {Dutta~Banik}, \citenamefont {Pandey}, \citenamefont {Majumdar},\ and\
  \citenamefont {Biswas}}]{DuttaBanik:2016jzv}%
  \BibitemOpen
  \bibfield  {author} {\bibinfo {author} {\bibfnamefont {A.}~\bibnamefont
  {Dutta~Banik}}, \bibinfo {author} {\bibfnamefont {M.}~\bibnamefont {Pandey}},
  \bibinfo {author} {\bibfnamefont {D.}~\bibnamefont {Majumdar}}, \ and\
  \bibinfo {author} {\bibfnamefont {A.}~\bibnamefont {Biswas}},\ }\href
  {\doibase 10.1140/epjc/s10052-017-5221-y} {\bibfield  {journal} {\bibinfo
  {journal} {Eur. Phys. J. C}\ }\textbf {\bibinfo {volume} {77}},\ \bibinfo
  {pages} {657} (\bibinfo {year} {2017})},\ \Eprint
  {http://arxiv.org/abs/1612.08621} {arXiv:1612.08621 [hep-ph]} \BibitemShut
  {NoStop}%
\bibitem [{\citenamefont {Bhattacharya}\ \emph {et~al.}(2022)\citenamefont
  {Bhattacharya}, \citenamefont {Chakraborti},\ and\ \citenamefont
  {Pradhan}}]{Bhattacharya:2021rwh}%
  \BibitemOpen
  \bibfield  {author} {\bibinfo {author} {\bibfnamefont {S.}~\bibnamefont
  {Bhattacharya}}, \bibinfo {author} {\bibfnamefont {S.}~\bibnamefont
  {Chakraborti}}, \ and\ \bibinfo {author} {\bibfnamefont {D.}~\bibnamefont
  {Pradhan}},\ }\href {\doibase 10.1007/JHEP07(2022)091} {\bibfield  {journal}
  {\bibinfo  {journal} {JHEP}\ }\textbf {\bibinfo {volume} {07}},\ \bibinfo
  {pages} {091} (\bibinfo {year} {2022})},\ \Eprint
  {http://arxiv.org/abs/2110.06985} {arXiv:2110.06985 [hep-ph]} \BibitemShut
  {NoStop}%
\bibitem [{\citenamefont {Costa}\ \emph {et~al.}(2022)\citenamefont {Costa},
  \citenamefont {Khan},\ and\ \citenamefont {Kim}}]{Costa:2022lpy}%
  \BibitemOpen
  \bibfield  {author} {\bibinfo {author} {\bibfnamefont {F.}~\bibnamefont
  {Costa}}, \bibinfo {author} {\bibfnamefont {S.}~\bibnamefont {Khan}}, \ and\
  \bibinfo {author} {\bibfnamefont {J.}~\bibnamefont {Kim}},\ }\href {\doibase
  10.1007/JHEP12(2022)165} {\bibfield  {journal} {\bibinfo  {journal} {JHEP}\
  }\textbf {\bibinfo {volume} {12}},\ \bibinfo {pages} {165} (\bibinfo {year}
  {2022})},\ \Eprint {http://arxiv.org/abs/2209.13653} {arXiv:2209.13653
  [hep-ph]} \BibitemShut {NoStop}%
\bibitem [{\citenamefont {Bélanger}\ \emph {et~al.}(2015)\citenamefont
  {Bélanger}, \citenamefont {Boudjema}, \citenamefont {Pukhov},\ and\
  \citenamefont {Semenov}}]{Belanger:2014vza}%
  \BibitemOpen
  \bibfield  {author} {\bibinfo {author} {\bibfnamefont {G.}~\bibnamefont
  {Bélanger}}, \bibinfo {author} {\bibfnamefont {F.}~\bibnamefont {Boudjema}},
  \bibinfo {author} {\bibfnamefont {A.}~\bibnamefont {Pukhov}}, \ and\ \bibinfo
  {author} {\bibfnamefont {A.}~\bibnamefont {Semenov}},\ }\href {\doibase
  10.1016/j.cpc.2015.03.003} {\bibfield  {journal} {\bibinfo  {journal}
  {Comput. Phys. Commun.}\ }\textbf {\bibinfo {volume} {192}},\ \bibinfo
  {pages} {322} (\bibinfo {year} {2015})},\ \Eprint
  {http://arxiv.org/abs/1407.6129} {arXiv:1407.6129 [hep-ph]} \BibitemShut
  {NoStop}%
\bibitem [{\citenamefont {Semenov}(2016)}]{Semenov:2014rea}%
  \BibitemOpen
  \bibfield  {author} {\bibinfo {author} {\bibfnamefont {A.}~\bibnamefont
  {Semenov}},\ }\href {\doibase 10.1016/j.cpc.2016.01.003} {\bibfield
  {journal} {\bibinfo  {journal} {Comput. Phys. Commun.}\ }\textbf {\bibinfo
  {volume} {201}},\ \bibinfo {pages} {167} (\bibinfo {year} {2016})},\ \Eprint
  {http://arxiv.org/abs/1412.5016} {arXiv:1412.5016 [physics.comp-ph]}
  \BibitemShut {NoStop}%
\bibitem [{\citenamefont {D'Eramo}\ and\ \citenamefont
  {Lenoci}(2021)}]{DEramo:2020gpr}%
  \BibitemOpen
  \bibfield  {author} {\bibinfo {author} {\bibfnamefont {F.}~\bibnamefont
  {D'Eramo}}\ and\ \bibinfo {author} {\bibfnamefont {A.}~\bibnamefont
  {Lenoci}},\ }\href {\doibase 10.1088/1475-7516/2021/10/045} {\bibfield
  {journal} {\bibinfo  {journal} {JCAP}\ }\textbf {\bibinfo {volume} {10}},\
  \bibinfo {pages} {045} (\bibinfo {year} {2021})},\ \Eprint
  {http://arxiv.org/abs/2012.01446} {arXiv:2012.01446 [hep-ph]} \BibitemShut
  {NoStop}%
\bibitem [{\citenamefont {Decant}\ \emph {et~al.}(2022)\citenamefont {Decant},
  \citenamefont {Heisig}, \citenamefont {Hooper},\ and\ \citenamefont
  {Lopez-Honorez}}]{Decant:2021mhj}%
  \BibitemOpen
  \bibfield  {author} {\bibinfo {author} {\bibfnamefont {Q.}~\bibnamefont
  {Decant}}, \bibinfo {author} {\bibfnamefont {J.}~\bibnamefont {Heisig}},
  \bibinfo {author} {\bibfnamefont {D.~C.}\ \bibnamefont {Hooper}}, \ and\
  \bibinfo {author} {\bibfnamefont {L.}~\bibnamefont {Lopez-Honorez}},\ }\href
  {\doibase 10.1088/1475-7516/2022/03/041} {\bibfield  {journal} {\bibinfo
  {journal} {JCAP}\ }\textbf {\bibinfo {volume} {03}},\ \bibinfo {pages} {041}
  (\bibinfo {year} {2022})},\ \Eprint {http://arxiv.org/abs/2111.09321}
  {arXiv:2111.09321 [astro-ph.CO]} \BibitemShut {NoStop}%
\bibitem [{\citenamefont {Feng}\ \emph
  {et~al.}(2003{\natexlab{a}})\citenamefont {Feng}, \citenamefont {Rajaraman},\
  and\ \citenamefont {Takayama}}]{Feng:2003uy}%
  \BibitemOpen
  \bibfield  {author} {\bibinfo {author} {\bibfnamefont {J.~L.}\ \bibnamefont
  {Feng}}, \bibinfo {author} {\bibfnamefont {A.}~\bibnamefont {Rajaraman}}, \
  and\ \bibinfo {author} {\bibfnamefont {F.}~\bibnamefont {Takayama}},\ }\href
  {\doibase 10.1103/PhysRevD.68.063504} {\bibfield  {journal} {\bibinfo
  {journal} {Phys. Rev. D}\ }\textbf {\bibinfo {volume} {68}},\ \bibinfo
  {pages} {063504} (\bibinfo {year} {2003}{\natexlab{a}})},\ \Eprint
  {http://arxiv.org/abs/hep-ph/0306024} {arXiv:hep-ph/0306024} \BibitemShut
  {NoStop}%
\bibitem [{\citenamefont {Feng}\ \emph
  {et~al.}(2003{\natexlab{b}})\citenamefont {Feng}, \citenamefont {Rajaraman},\
  and\ \citenamefont {Takayama}}]{Feng:2003xh}%
  \BibitemOpen
  \bibfield  {author} {\bibinfo {author} {\bibfnamefont {J.~L.}\ \bibnamefont
  {Feng}}, \bibinfo {author} {\bibfnamefont {A.}~\bibnamefont {Rajaraman}}, \
  and\ \bibinfo {author} {\bibfnamefont {F.}~\bibnamefont {Takayama}},\ }\href
  {\doibase 10.1103/PhysRevLett.91.011302} {\bibfield  {journal} {\bibinfo
  {journal} {Phys. Rev. Lett.}\ }\textbf {\bibinfo {volume} {91}},\ \bibinfo
  {pages} {011302} (\bibinfo {year} {2003}{\natexlab{b}})},\ \Eprint
  {http://arxiv.org/abs/hep-ph/0302215} {arXiv:hep-ph/0302215} \BibitemShut
  {NoStop}%
\bibitem [{\citenamefont {Yaguna}\ and\ \citenamefont
  {Zapata}(2020)}]{Yaguna:2019cvp}%
  \BibitemOpen
  \bibfield  {author} {\bibinfo {author} {\bibfnamefont {C.~E.}\ \bibnamefont
  {Yaguna}}\ and\ \bibinfo {author} {\bibfnamefont {O.}~\bibnamefont
  {Zapata}},\ }\href {\doibase 10.1007/JHEP03(2020)109} {\bibfield  {journal}
  {\bibinfo  {journal} {JHEP}\ }\textbf {\bibinfo {volume} {03}},\ \bibinfo
  {pages} {109} (\bibinfo {year} {2020})},\ \Eprint
  {http://arxiv.org/abs/1911.05515} {arXiv:1911.05515 [hep-ph]} \BibitemShut
  {NoStop}%
\bibitem [{\citenamefont {Yaguna}\ and\ \citenamefont
  {Zapata}(2021)}]{Yaguna:2021vhb}%
  \BibitemOpen
  \bibfield  {author} {\bibinfo {author} {\bibfnamefont {C.~E.}\ \bibnamefont
  {Yaguna}}\ and\ \bibinfo {author} {\bibfnamefont {O.}~\bibnamefont
  {Zapata}},\ }\href {\doibase 10.1007/JHEP10(2021)185} {\bibfield  {journal}
  {\bibinfo  {journal} {JHEP}\ }\textbf {\bibinfo {volume} {10}},\ \bibinfo
  {pages} {185} (\bibinfo {year} {2021})},\ \Eprint
  {http://arxiv.org/abs/2106.11889} {arXiv:2106.11889 [hep-ph]} \BibitemShut
  {NoStop}%
\bibitem [{\citenamefont {Batell}(2011)}]{Batell:2010bp}%
  \BibitemOpen
  \bibfield  {author} {\bibinfo {author} {\bibfnamefont {B.}~\bibnamefont
  {Batell}},\ }\href {\doibase 10.1103/PhysRevD.83.035006} {\bibfield
  {journal} {\bibinfo  {journal} {Phys. Rev.}\ }\textbf {\bibinfo {volume}
  {D83}},\ \bibinfo {pages} {035006} (\bibinfo {year} {2011})},\ \Eprint
  {http://arxiv.org/abs/1007.0045} {arXiv:1007.0045 [hep-ph]} \BibitemShut
  {NoStop}%
\bibitem [{\citenamefont {Belanger}\ \emph {et~al.}(2012)\citenamefont
  {Belanger}, \citenamefont {Kannike}, \citenamefont {Pukhov},\ and\
  \citenamefont {Raidal}}]{Belanger:2012vp}%
  \BibitemOpen
  \bibfield  {author} {\bibinfo {author} {\bibfnamefont {G.}~\bibnamefont
  {Belanger}}, \bibinfo {author} {\bibfnamefont {K.}~\bibnamefont {Kannike}},
  \bibinfo {author} {\bibfnamefont {A.}~\bibnamefont {Pukhov}}, \ and\ \bibinfo
  {author} {\bibfnamefont {M.}~\bibnamefont {Raidal}},\ }\href {\doibase
  10.1088/1475-7516/2012/04/010} {\bibfield  {journal} {\bibinfo  {journal}
  {JCAP}\ }\textbf {\bibinfo {volume} {1204}},\ \bibinfo {pages} {010}
  (\bibinfo {year} {2012})},\ \Eprint {http://arxiv.org/abs/1202.2962}
  {arXiv:1202.2962 [hep-ph]} \BibitemShut {NoStop}%
\bibitem [{\citenamefont {Bélanger}\ \emph {et~al.}(2014)\citenamefont
  {Bélanger}, \citenamefont {Kannike}, \citenamefont {Pukhov},\ and\
  \citenamefont {Raidal}}]{Belanger:2014bga}%
  \BibitemOpen
  \bibfield  {author} {\bibinfo {author} {\bibfnamefont {G.}~\bibnamefont
  {Bélanger}}, \bibinfo {author} {\bibfnamefont {K.}~\bibnamefont {Kannike}},
  \bibinfo {author} {\bibfnamefont {A.}~\bibnamefont {Pukhov}}, \ and\ \bibinfo
  {author} {\bibfnamefont {M.}~\bibnamefont {Raidal}},\ }\href {\doibase
  10.1088/1475-7516/2014/06/021} {\bibfield  {journal} {\bibinfo  {journal}
  {JCAP}\ }\textbf {\bibinfo {volume} {1406}},\ \bibinfo {pages} {021}
  (\bibinfo {year} {2014})},\ \Eprint {http://arxiv.org/abs/1403.4960}
  {arXiv:1403.4960 [hep-ph]} \BibitemShut {NoStop}%
\bibitem [{\citenamefont {Nurmi}\ \emph {et~al.}(2015)\citenamefont {Nurmi},
  \citenamefont {Tenkanen},\ and\ \citenamefont {Tuominen}}]{Nurmi:2015ema}%
  \BibitemOpen
  \bibfield  {author} {\bibinfo {author} {\bibfnamefont {S.}~\bibnamefont
  {Nurmi}}, \bibinfo {author} {\bibfnamefont {T.}~\bibnamefont {Tenkanen}}, \
  and\ \bibinfo {author} {\bibfnamefont {K.}~\bibnamefont {Tuominen}},\ }\href
  {\doibase 10.1088/1475-7516/2015/11/001} {\bibfield  {journal} {\bibinfo
  {journal} {JCAP}\ }\textbf {\bibinfo {volume} {11}},\ \bibinfo {pages} {001}
  (\bibinfo {year} {2015})},\ \Eprint {http://arxiv.org/abs/1506.04048}
  {arXiv:1506.04048 [astro-ph.CO]} \BibitemShut {NoStop}%
\bibitem [{\citenamefont {Kainulainen}\ \emph {et~al.}(2016)\citenamefont
  {Kainulainen}, \citenamefont {Nurmi}, \citenamefont {Tenkanen}, \citenamefont
  {Tuominen},\ and\ \citenamefont {Vaskonen}}]{Kainulainen:2016vzv}%
  \BibitemOpen
  \bibfield  {author} {\bibinfo {author} {\bibfnamefont {K.}~\bibnamefont
  {Kainulainen}}, \bibinfo {author} {\bibfnamefont {S.}~\bibnamefont {Nurmi}},
  \bibinfo {author} {\bibfnamefont {T.}~\bibnamefont {Tenkanen}}, \bibinfo
  {author} {\bibfnamefont {K.}~\bibnamefont {Tuominen}}, \ and\ \bibinfo
  {author} {\bibfnamefont {V.}~\bibnamefont {Vaskonen}},\ }\href {\doibase
  10.1088/1475-7516/2016/06/022} {\bibfield  {journal} {\bibinfo  {journal}
  {JCAP}\ }\textbf {\bibinfo {volume} {06}},\ \bibinfo {pages} {022} (\bibinfo
  {year} {2016})},\ \Eprint {http://arxiv.org/abs/1601.07733} {arXiv:1601.07733
  [astro-ph.CO]} \BibitemShut {NoStop}%
\bibitem [{\citenamefont {Bernal}\ and\ \citenamefont
  {Xu}(2022)}]{Bernal:2022wck}%
  \BibitemOpen
  \bibfield  {author} {\bibinfo {author} {\bibfnamefont {N.}~\bibnamefont
  {Bernal}}\ and\ \bibinfo {author} {\bibfnamefont {Y.}~\bibnamefont {Xu}},\
  }\href {\doibase 10.1088/1475-7516/2022/12/017} {\bibfield  {journal}
  {\bibinfo  {journal} {JCAP}\ }\textbf {\bibinfo {volume} {12}},\ \bibinfo
  {pages} {017} (\bibinfo {year} {2022})},\ \Eprint
  {http://arxiv.org/abs/2209.07546} {arXiv:2209.07546 [hep-ph]} \BibitemShut
  {NoStop}%
\bibitem [{\citenamefont {Silva-Malpartida}\ \emph {et~al.}(2023)\citenamefont
  {Silva-Malpartida}, \citenamefont {Bernal}, \citenamefont {Jones-P\'erez},\
  and\ \citenamefont {Lineros}}]{Silva-Malpartida:2023yks}%
  \BibitemOpen
  \bibfield  {author} {\bibinfo {author} {\bibfnamefont {J.}~\bibnamefont
  {Silva-Malpartida}}, \bibinfo {author} {\bibfnamefont {N.}~\bibnamefont
  {Bernal}}, \bibinfo {author} {\bibfnamefont {J.}~\bibnamefont
  {Jones-P\'erez}}, \ and\ \bibinfo {author} {\bibfnamefont {R.~A.}\
  \bibnamefont {Lineros}},\ }\href@noop {} {\  (\bibinfo {year} {2023})},\
  \Eprint {http://arxiv.org/abs/2306.14943} {arXiv:2306.14943 [hep-ph]}
  \BibitemShut {NoStop}%
\bibitem [{\citenamefont {Becker}\ \emph {et~al.}(2023)\citenamefont {Becker},
  \citenamefont {Copello}, \citenamefont {Harz}, \citenamefont {Lang},\ and\
  \citenamefont {Xu}}]{Becker:2023tvd}%
  \BibitemOpen
  \bibfield  {author} {\bibinfo {author} {\bibfnamefont {M.}~\bibnamefont
  {Becker}}, \bibinfo {author} {\bibfnamefont {E.}~\bibnamefont {Copello}},
  \bibinfo {author} {\bibfnamefont {J.}~\bibnamefont {Harz}}, \bibinfo {author}
  {\bibfnamefont {J.}~\bibnamefont {Lang}}, \ and\ \bibinfo {author}
  {\bibfnamefont {Y.}~\bibnamefont {Xu}},\ }\href@noop {} {\  (\bibinfo {year}
  {2023})},\ \Eprint {http://arxiv.org/abs/2306.17238} {arXiv:2306.17238
  [hep-ph]} \BibitemShut {NoStop}%
\bibitem [{\citenamefont {Tenkanen}(2016)}]{Tenkanen:2016twd}%
  \BibitemOpen
  \bibfield  {author} {\bibinfo {author} {\bibfnamefont {T.}~\bibnamefont
  {Tenkanen}},\ }\href {\doibase 10.1007/JHEP09(2016)049} {\bibfield  {journal}
  {\bibinfo  {journal} {JHEP}\ }\textbf {\bibinfo {volume} {09}},\ \bibinfo
  {pages} {049} (\bibinfo {year} {2016})},\ \Eprint
  {http://arxiv.org/abs/1607.01379} {arXiv:1607.01379 [hep-ph]} \BibitemShut
  {NoStop}%
\bibitem [{\citenamefont {Almeida}\ \emph {et~al.}(2019)\citenamefont
  {Almeida}, \citenamefont {Bernal}, \citenamefont {Rubio},\ and\ \citenamefont
  {Tenkanen}}]{Almeida:2018oid}%
  \BibitemOpen
  \bibfield  {author} {\bibinfo {author} {\bibfnamefont {J.~P.~B.}\
  \bibnamefont {Almeida}}, \bibinfo {author} {\bibfnamefont {N.}~\bibnamefont
  {Bernal}}, \bibinfo {author} {\bibfnamefont {J.}~\bibnamefont {Rubio}}, \
  and\ \bibinfo {author} {\bibfnamefont {T.}~\bibnamefont {Tenkanen}},\ }\href
  {\doibase 10.1088/1475-7516/2019/03/012} {\bibfield  {journal} {\bibinfo
  {journal} {JCAP}\ }\textbf {\bibinfo {volume} {03}},\ \bibinfo {pages} {012}
  (\bibinfo {year} {2019})},\ \Eprint {http://arxiv.org/abs/1811.09640}
  {arXiv:1811.09640 [hep-ph]} \BibitemShut {NoStop}%
\end{thebibliography}%

\end{document}